\documentclass[sigconf]{acmart}
\usepackage{soul}
\usepackage{multirow}

\AtBeginDocument{%
  }

\copyrightyear{2025} 
\acmYear{2025} 
\setcopyright{cc}
\setcctype{by}
\acmConference[CHI '25]{CHI Conference on Human Factors in Computing Systems}{April 26-May 1, 2025}{Yokohama, Japan}
\acmBooktitle{CHI Conference on Human Factors in Computing Systems (CHI '25), April 26-May 1, 2025, Yokohama, Japan}
\acmDOI{10.1145/3706598.3713971} 
\acmISBN{979-8-4007-1394-1/25/04}

\begin{document}

\title{Co-designing Large Language Model Tools for Project-Based Learning with K-12 Educators}


\author{Prerna Ravi}
\affiliation{%
  \institution{Massachusetts Institute of Technology}
  \city{Cambridge, MA}
  \country{USA}}
\email{prernar@mit.edu}

\author{John Masla}
\affiliation{%
  \institution{Massachusetts Institute of Technology}
  \city{Cambridge, MA}
  \country{USA}}
\email{j_masla@mit.edu}

\author{Gisella Kakoti}
\affiliation{%
  \institution{Massachusetts Institute of Technology}
  \city{Cambridge, MA}
  \country{USA}}
\email{gkakoti@alum.mit.edu}

\author{Grace C. Lin}
\affiliation{%
  \institution{Massachusetts Institute of Technology}
  \city{Cambridge, MA}
  \country{USA}}
\email{gcl@mit.edu}

\author{Emma Anderson}
\affiliation{%
  \institution{Massachusetts Institute of Technology}
  \city{Cambridge, MA}
  \country{USA}}
\email{eanderso@mit.edu}

\author{Matt Taylor}
\affiliation{%
  \institution{Massachusetts Institute of Technology}
  \city{Cambridge, MA}
  \country{USA}}
\email{mewtaylor@gmail.com}

\author{Anastasia K. Ostrowski}
\affiliation{%
  \institution{Purdue University}
  \city{West Lafayette, IN}
  \country{USA}}
\email{akostrow@purdue.edu}

\author{Cynthia Breazeal}
\affiliation{%
  \institution{Massachusetts Institute of Technology}
  \city{Cambridge MA}
  \country{USA}}
\email{cynthiab@media.mit.edu}

\author{Eric Klopfer}
\affiliation{%
  \institution{Massachusetts Institute of Technology}
  \city{Cambridge MA}
  \country{USA}}
\email{klopfer@mit.edu}

\author{Hal Abelson}
\affiliation{%
  \institution{Massachusetts Institute of Technology}
  \city{Cambridge MA}
  \country{USA}}
\email{hal@mit.edu}







\renewcommand{\shortauthors}{Ravi, et al.}

\begin{abstract}
  The emergence of generative AI, particularly large language models (LLMs), has opened the door for student-centered and active learning methods like project-based learning (PBL). However, PBL poses practical implementation challenges for educators around project design and management, assessment, and balancing student guidance with student autonomy. The following research documents a co-design process with interdisciplinary K-12 teachers to explore and address the current PBL challenges they face. Through teacher-driven interviews, collaborative workshops, and iterative design of wireframes, we gathered evidence for ways LLMs can support teachers in implementing high-quality PBL pedagogy by automating routine tasks and enhancing personalized learning. Teachers in the study advocated for supporting their professional growth and augmenting their current roles without replacing them. They also identified affordances and challenges around classroom integration, including resource requirements and constraints, ethical concerns, and potential immediate and long-term impacts. Drawing on these, we propose design guidelines for future deployment of LLM tools in PBL.
\end{abstract}




\maketitle

\section{INTRODUCTION}
Project-based learning (PBL) has gained prominence as a K-12 educational approach that immerses students in meaningful, real-world tasks, fostering deeper learning experiences \cite{boaler1998open, panasan2010learning, schneider2002performance, chen2019revisiting, PBLWorks}. 
Unlike traditional instructional methods, PBL emphasizes student-centered pedagogy, where learners actively construct knowledge through exploration, collaboration, and reflection \cite{condliffe2017project, noguera2015equal, peterson2012uncovering}. 
By providing students with opportunities to explore topics of personal and cultural relevance, PBL supports diverse learning needs and backgrounds, making education more inclusive and accessible.

While PBL offers significant benefits to students, it also necessitates a fundamental shift in educators' roles from knowledge providers to learning facilitators \cite{ertmer2006jumping, savery2015overview}. This requires educators to adopt new pedagogical strategies, manage complex classroom dynamics and provide ongoing, formative assessments that support student learning \cite{sahoo2013formative, albanese2019types}.
Challenges to PBL implementation include designing standards-aligned projects that meet diverse student needs, managing planning time, transitioning students to active roles \cite{simons2004instructional, chen2019revisiting, thomas1998project, zheng2024charting, viro2020teachers, ertmer2006jumping, gallagher1997problem},  creating authentic driving questions and assessing interpersonal skills, which traditional assessment methods often fail to capture \cite{markula2022key, blumenfeld1991motivating, mentzer2017examination, zheng2024charting, aksela2019project, colley2006understanding, brinkerhoff2004support}. Such challenges emphasize the need for nuanced instructional strategies for PBL that better capture the scope of student learning \cite{zheng2024charting}.


To address these challenges, the Human-Computer Interaction (HCI) community has shown a rising interest in generative artificial intelligence (GenAI) systems for education \cite{muller2022genaichi, sun2024generative, weisz2024design, han2024teachers, zheng2024charting, cai2024advancing, shaer2024ai}. 
Large language models (LLMs) in particular 
are uniquely suited to
address the inherent complexities of PBL {\cite{gustafson2025enhancing, zheng2024selfgauge, nikolicsupporting}} due to their capacity to personalize learning and offer real-time custom feedback with the potential to support differentiated instruction in PBL contexts \cite{asrifan2024integrating, kong2024developing, alam2023intelligence}. 
By automating routine tasks and augmenting project workflows, these tools could help streamline classroom management, thereby supporting teachers in delivering more engaging learning experiences \cite{zheng2024charting, krajcik2006project}.
Given the potential of LLMs to transform PBL, it is essential that these tools are designed prioritizing educators' needs and experiences, with teachers actively involved in the design process \cite{penuel2007designing, bakah2012updating, wan2021exploratory, handelzalts2019collaborative, mckenney2016collaborative}. 

\subsection{Our Contribution}

To the best of our knowledge, this is the first co-design study with K-12 teachers aimed specifically at incorporating LLMs into PBL pedagogy.  Given the unprecedented and rapidly advancing nature of GenAI and its ethical considerations, our research is particularly timely. Our research objectives (RO) include:

\begin{enumerate}
    \item \textbf{Explore Demands and Challenges in PBL:} We investigated specific challenges educators encounter when implementing PBL in their classrooms, focusing on how they assess student learning outcomes in areas such as teamwork, creativity, and problem-solving. 
    \item \textbf{Co-design GenAI Tools for PBL:} Next, we explored how GenAI-powered tools, co-designed with interdisciplinary educators, could address diverse needs and challenges of PBL environments. Our co-design workshops examined how these tools could alleviate teachers' workloads while supporting their current practices, pedagogical goals, and professional growth. We constructed wireframes embodying teachers' values 
    and received iterative feedback, {from teachers with diverse expertise,} on their potential integration into PBL environments.
    \item \textbf{Outline Design Considerations for LLM Tools in PBL:} Finally, we propose a set of design guidelines for LLM tools that educational technology designers and educators may find valuable for assessing student progress, setting learning goals, providing iterative and personalized feedback, and managing resources, timelines, and artifacts across different student projects and milestones.
\end{enumerate}

Our findings provide valuable insights into shaping the future use of LLMs in PBL, \textit{guiding and encouraging} its responsible implementation in classrooms. 
{It should be noted that although we used the term GenAI for the study framing with our participants, they focused on a particular type of GenAI: LLMs for their proposed solutions. Because of this duality, both these terms are used in this paper.}

\section{RELATED WORK}


\subsection{Project-Based Learning}

Rooted in constructivist learning theories, project-based learning (PBL) places students at the center of the learning process, challenging conventional education practices and redefining the roles of teaching, learning, and school organization \cite{fitzgerald2020overlapping, guo2023effects, kokotsaki2016project, woods2024}.  
Unlike problem-based learning, which promotes promotes deductive reasoning \cite{condliffe2017project, thomas1999project}, and involves structured evaluation based on how well a student’s solution addresses a defined problem (e.g., designing a more effective trash can to improve adoption or conducting participatory action research on a social issue), project-based learning engages students in complex, real-world tasks to create a broader range of artifacts (e.g. reports, models, and presentations) that demonstrate learning \cite{krajcik2006project, oguz2014comparison, thomas2010review, chen2019revisiting, barron2008teaching, thomas1999project} (e.g., using photojournalism to explore local flora and fauna or writing a comic book to analyze tropes in the hero’s journey). 
This fosters creativity and critical thinking through open-ended driving questions (e.g. “what is the proper role of government in a democracy?”) \cite{parker2011rethinking, blumenfeld1991motivating, diehl1999project, svihla2020facilitating}.
Research has consistently demonstrated the numerous benefits of PBL across various educational contexts. Unlike traditional instruction, PBL (1) encourages students to delve deeply into the subject matter, fostering a more profound understanding of the material \cite{boaler1998open, panasan2010learning, schneider2002performance, chen2019revisiting}; (2) enhances mastery of subjects and supports the development of critical 21st-century skills like collaboration and communication \cite{condliffe2017project, noguera2015equal, peterson2012uncovering, chen2019revisiting}; (3) boosts motivation, and engagement \cite{hernandez2009learning,kaldi2011project, blumenfeld1991motivating, holm2011project, bender2012project, intel2007designing}; 
and (4) develops metacognitive skills by helping students assess their progress and continuously improve \cite{thomas1998project, thomas1999project, english2013supporting, bender2012project, tseng2013attitudes}. 

While PBL is inherently student-centered, the success of this approach is heavily influenced by the role of teachers as facilitators, guiding and scaffolding student learning
\cite{barron2008teaching}. This shift in roles and professional identity requires teachers to adapt their pedagogical approaches and develop new skills, often leading to greater ownership, self-efficacy, and confidence in their teaching \cite{choi2019does, havice2018evaluating, potvin2021consequential}. However, transitioning to PBL often requires targeted professional development (PD) that provides teachers with peer collaboration, reflection, and ongoing support during PBL implementation \cite{Aitken2019, dunbar2022shifting, blumenfeld1991motivating, quint2018project, park2018equip}. 
Given the pivotal role of teachers in harnessing PBL benefits, our research reveals their needs and explores how LLMs could support them in successfully implementing PBL and foster continued professional growth. 


To maximize the effectiveness of these tools, it is essential to consider challenges teachers encounter during PBL implementation, especially around generating authentic driving questions, 
managing time and group work, and balancing instructor-led guidance with student-directed learning \cite{zheng2024charting, thomas2010review, mergendoller2005managing}. 
Students’ discomfort with the cognitive and social demands of PBL can lead to frustration, particularly among high-achieving students accustomed to traditional instruction \cite{condliffe2017project}. In addition, teachers face challenges initiating inquiry, facilitating dialogue, and scaffolding learning \cite{reiser2006making, kali2008technology, krajcik2014promises, quintana2018scaffolding, reiser2018scaffolding}. 
Assessing student learning in PBL is also notably difficult, as traditional tests often fail to capture the depth of understanding that PBL aims to develop.  
Performance-based assessments are hard to implement reliably \cite{hertzog2007transporting, mergendoller2005managing, darling2010beyond, aslan2015examining}, student artifacts can be difficult to score consistently, and teachers lack the time to provide personalized student feedback \cite{hattie2011instruction}. 
Integrating technology is also challenging due to broader institutional factors such as limited resources, district mandates, and lack of school tech maintenance and support \cite{zheng2024charting}. Our study unpacks these PBL challenges in the literature through the nuanced experiences of teachers in interdisciplinary settings and explores how LLMs impact their roles and instructional practices.

\subsection{AI Tools to Support PBL Pedagogical Needs}

AI's capacity to personalize learning is beneficial in PBL settings where differentiated instruction is crucial. AI tools can support targeted instruction by tailoring lessons for diverse learners, encouraging student iteration on artifacts, and adapting materials to individual student strengths and weaknesses \cite{asrifan2024integrating, kong2024developing, alam2023intelligence}. 
Helping students manage time and materials efficiently, tools like Trello and Cronofy, integrated with AI plug-ins, can predict resource needs  \cite{tanga2024exploration}. 
AI-assisted design tools like Autodesk Dreamcatcher and intelligent project management software such as Asana and Monday.com can facilitate student project management, allowing them to focus on creative and critical thinking \cite{tanga2024exploration}.

Considering project collaboration and knowledge construction\cite{tanga2024exploration, dutta2024enhancing, asrifan2024integrating}, research scholars have discussed how AI can evaluate group performance by analyzing interaction patterns and predicting outcomes based on academic and behavioral data, enabling more effective grouping strategies \cite{cen2016quantitative, schneider2014toward}. AI systems can also guide students toward productive problem-solving, suggesting activities on collaborative styles, and pedagogical interventions to improve group work \cite{adamson2014towards, dyke2013enhancing, kumar2010architecture}. 
Considering grading in PBL, AI-powered assessment systems, such as Automated Essay Scoring (AES) and Automated Written Corrective Feedback (AWCF), can offer real-time, continuous feedback, helping students to refine their work iteratively \cite{rudolph2023chatgpt, cope2021artificial}. These systems not only reduce the teacher's workload but also enhance the accuracy and efficiency of grading, allowing teachers to focus on more meaningful interactions with students \cite{owan2023exploring}. AI can also support formative assessment practices in PBL by providing insights into students' progress, helping teachers guide and support students more effectively \cite{lan2024teachers}.

Despite potential benefits, integrating LLMs into PBL presents challenges, such as biased algorithms, that necessitates careful design and human oversight to ensure fairness and accuracy in assessment and feedback \cite{schneider2023towards}. Comprehensive PD is needed for teachers to use LLM tools responsibly and equitably \cite{lan2024teachers, askarbekuly2024llm}, understand
their long-term effects across age groups and subject areas, and address ethical concerns like data privacy \cite{asrifan2024integrating, wang2024artificial, zha2024designing}. In our current work, we aim to overcome some of these challenges by co-designing GenAI-powered support systems with \textit{teachers} in PBL settings. {
This is unlike recent studies that have focused on co-designing these tools with \textit{students} in the higher education contexts \cite{zheng2024selfgauge, zheng2024charting, nikolicsupporting, gustafson2025enhancing}. We also examine challenges more unique to K-12 PBL, that emphasizes scaffolding and skill-building \cite{condliffe2017project}.} 

\subsection{Co-designing Educational Tools with Teachers}

Co-design is a collaborative, iterative process where teachers, researchers, and developers jointly design, prototype, and evaluate educational tools 
\cite{roschelle2006co}. This approach leverages stakeholders' expertise to address concrete educational needs, making it well-suited for creating technology-enhanced learning environments that are contextually relevant and practical for real-world classrooms \cite{cober2015teachers, brown1992design, lingnau2007empowering}. Studies have documented the co-design of mobile science applications, scripted wiki environments, and other educational technologies that support authentic scientific inquiry and other pedagogical practices \cite{spikol2009integrating, zhang2010deconstructing, peters2009co, cober2015teachers}.

Involving teachers in the co-design process enhances development efficiency, increases teacher agency and ownership, and promotes higher adoption rates and sustained use of educational tools beyond the initial study \cite{penuel2007designing, bakah2012updating, wan2021exploratory, handelzalts2019collaborative, mckenney2016collaborative, lin2021engaging}. 
Co-design can also serve as a form of professional development, enabling teachers to deepen their understanding of new technologies and explore ways to incorporate them into instructional strategies \cite{wan2021exploratory, bakah2012updating, voogt2015collaborative}. Because co-design fosters a more reciprocal and participatory approach {\cite{alfredo2024human}}, it has the potential to upend the traditional, unidirectional process of educational technology  \cite{teeters2016challenge} and enables teachers to engage with complex topics beyond their expertise \cite{disalvo2017participatory}. 
Furthermore, co-design has the potential to influence the broader sociocultural context of schools and drive systemic change in educational technology design and implementation by reshaping power dynamics and promoting an ethics of care within educational research and design \cite{wake2013developing, higgins2019power, matuk2021students}. {Scholars have however noted that involving teachers in the human centered design process happens less often in K-12 settings compared to higher education \cite{topali2024designing}.}



\section{METHODS}

We conducted two studies: the first with expert PBL teachers and the second with teachers {who had varying levels of PBL expertise. We explain our definition for expertise in section 3.1.1.} 
By involving both groups, we sought to 
address the varied challenges and opportunities in PBL classrooms.  {Our studies were approved by the Institutional Review Board (IRB) at Massachusetts Institute of Technology. We obtained informed consent from all participants.}

\subsection{STUDY 1} 
In this study, we engaged interdisciplinary {teachers with high levels of PBL experience and comfort. We worked with experts to leverage their deep pedagogical knowledge and refined understanding of effective PBL strategies, accumulated through years of trial, reflection, and adaptation.
They participated} in semi-structured interviews, a two-step co-design process, and feedback sessions on wireframes. We followed guidelines put forth in the literature \cite{ostrowski2021long, liu2003towards} by incorporating a mix of both divergent and convergent design thinking opportunities for participants. The divergent stages facilitated the generation of numerous ideas and concepts, while convergent stages focused on refining and narrowing down these ideas \cite{ostrowski2021long}. By offering diverse tools and activities, we enabled participants to discover the most effective methods for articulating their thoughts and generating new ideas \cite{ostrowski2021long}.

\subsubsection{\textbf{Participants and Recruitment}}

Eleven participants were recruited via mailing lists, direct emails, collaborations, and social media across the United States. {We specifically recruited participants with high levels of experience and comfort with PBL. This was the only precondition to participate in the study.}
Teachers were compensated \$200 for participation (of approximately 10 hours) on this study.

Teachers completed a pre-survey covering demographic information and details regarding their past PBL and GenAI experience as part of the recruitment process. {We received over 50 responses from interested teachers.}
We {narrowed this list down to select our final 11 participants for the study \cite{caine2016local}. We aimed to have a diverse sample by selecting teachers} from multiple U.S. states, school systems (rural/urban/suburban areas as well as public/private schools with varied resource access \cite{goh2016urban}), subject backgrounds, varying levels of GenAI experience, and diverse student age groups.
We also sought teachers from both STEM and non-STEM classrooms, those who work with historically marginalized students, and those who have taught PBL online \cite{eubanks2018automating, vanderlinde2010gap, ravi2021pandemic, trust2021emergency}.
We wanted to intentionally design GenAI PBL tools with {the above factors in mind from the beginning to ensure inclusive educational practices and} avoid exacerbating existing disparities \cite{eubanks2018automating}.

\begin{table*}[ht]
    \centering
    \begin{tabular}{|p{0.1\linewidth}|l |p{0.12\linewidth}|p{0.1\linewidth}|p{0.12\linewidth}|p{0.1\linewidth}|p{0.1\linewidth}|p{0.1\linewidth}|}
    \hline 
         \textbf{Participant}  & \textbf{Gender}   & \textbf{{Available School Details}}&  \textbf{Subjects} &  \textbf{Student Age-groups} &  \textbf{Years of experience with PBL} &  \textbf{Comfort levels with PBL} & \textbf{GenAI Tool Experience}\\ \hline \hline 
         P01 &F  &public, urban, socioeconomically disadvantaged students&  English&  9th-12th grade, higher ed, adults&  More than 5 years&  5& 5\\ \hline 
         P02 &M  &public, suburban&  Computer Science&  9th-12th &  More than 5 years&  5& 3\\ \hline 
         P03 &M  &public, experiential learning school&  History&  9th-12th &  1-2 years&  4& 2\\ \hline 
         P04 &M  &public, urban, socioeconomically disadvantaged students&  Mathematics&  9th-12th &  3-5 years&  4& 3\\ \hline 
         P05 &F  &public, online teaching&  Design, Tech, and Eng&  9th-12th, higher ed, adults&  More than 5 years&  5& 4\\ \hline 
         P06 &F  &public, rural&  Design, Tech, and Eng&  3rd-5th, 6th-8th&  More than 5 years&  4& 5\\ \hline 
         P07 &M  &online teaching&  Mathematics&  3rd-5th grade, 6th-8th, 9th-12th&  More than 5 years&  5& 4\\ \hline 
         P08 &F  &private&  Computer Science&  3rd-5th, 6th-8th, 9th-12th, higher ed, adults&  3-5 years&  4& 3\\ \hline 
         P09 &M  &private&  History&  kindergarten, 1st-2nd, 3rd-5th, 6th-8th&  More than 5 years&  5& 4\\ \hline
 P10 &F  &private, urban& Design, Tech, and Eng& 9th-12th& More than 5 years& 5&4\\\hline
 P11 &F  &public, suburban& Computer Science& 9th-12th& More than 5 years& 5&2\\\hline
    \end{tabular}
    \caption{Study 1 participants' details, N=11}
    \label{tab:participants}
    \Description{The table provides details of the participants from Study 1, which includes teachers with high levels of expertise in project-based learning (PBL). The table lists 11 participants (P01 to P11) and includes information on their gender, school details, subjects taught, student age groups, years of experience with PBL, comfort levels with PBL, and experience with generative AI (GenAI) tools.
    Among the participants, there is a mix of male (M) and female (F) teachers across different subjects such as English, Computer Science, History, Math, and Design, Technology, and Engineering. They teach in a mix of private/ public and rural/urban/suburban schools. P05 and P07 teach in online settings, P01 and P04 work with socio-economically disadvantaged students, and P03 is a part of an experiential learning school.
    The teachers work with a wide range of student groups, including kindergarten, elementary (3rd-5th grade), middle school (6th-8th grade), high school (9th-12th grade), higher education, and adults. Most participants have more than five years of experience with PBL, with a few having between one to five years. Comfort levels with PBL range from 4 to 5 on a scale where 5 indicates the highest comfort. Similarly, their experience with GenAI tools varies, with scores ranging from 2 to 5, reflecting varying levels of familiarity and usage.}
\end{table*}

Table \ref{tab:participants} breaks down the participants recruited (from the pre-survey responses). The teachers instructed various subjects, including computer science (n = 3), history (n = 2), English (n = 1), math (n = 2), and design, tech, and engineering (n = 3). Additionally, P06 served as a tech integration specialist, collaborating with other teachers to design and implement technology-enhanced projects. P05 and P07 taught PBL in online settings.
{Teachers,  on average, had 5+ years of experience implementing PBL and reported an average comfort level of 4.6/5. In this study, we define PBL expertise as a combination of years of experience and comfort levels (showing self-efficacy) resulting from focused PBL implementation \cite{rand}. Comfort, in particular, is influenced not only by prior experience but also by the school systems teachers operate within. For instance, P03 described themselves as a PBL expert despite having relatively few years of experience, attributing their confidence to the rigorous training and practice received from working at a school strongly focused on experiential learning \cite{lam2010school}.}

\subsubsection{\textbf{Data Collection}}
\paragraph{\textbf{Interviews}}

Between the months of March and April 2024, we conducted 11 semi-structured interviews \cite{kvale2009interviews}. All interviews were audio recorded with participant consent, ranged from 60-90 minutes each, and were conducted remotely via Zoom. We used these interviews as an opportunity to build rapport with our participants prior to co-design. The narrative style of the interview encouraged participants to engage in divergent thinking, allowing them to respond to open-ended questions freely with minimal redirection from the interviewers \cite{ostrowski2021long}.

During the interviews, we asked participants about their teaching contexts and experiences with PBL, starting with its integration into the curriculum, alignment with other methods, technology use, and contextual factors like teacher-student ratios. We then explored strategies for maintaining coherence across projects, fostering student engagement, and tracking progress. A significant portion of the interviews focused on assessments, including approaches to grading, rubric use, and challenges with differentiation and progress tracking. We also asked teachers about helping students transition from traditional learning to PBL, and their views on scaffolding techniques and the open-ended nature of PBL.
The recruitment survey responses from our 11 participants informed these interview questions (see Table \ref{tab:participants}). In addition, the broader topics in the protocol were decided based on the discussions outlined in the literature review by Condliffe et al. \cite{condliffe2017project}.

\paragraph{\textbf{Co-design Workshops}}

We conducted two co-design workshops (each lasting 2.5-3 hours) with teachers to explore how GenAI tools could support diverse PBL classroom needs without disrupting current practices. 
Due to teachers' limited availability and varying schedules, each workshop was divided into three smaller group sessions with three to four teachers, enabling focused, interactive discussions in a manageable online setting. The sessions, which included teachers from different subject areas to ensure diverse perspectives, were conducted remotely via Zoom and recorded with participants' consent. Below, we outline the structure and activities of each workshop:

\paragraph{\textbf{Workshop 1}}

The first co-design workshop aimed to identify key PBL challenges and explore how GenAI could help alleviate teacher workload. Each session included three to four teachers, a facilitator, and a notetaker. The make-up of each group was:

\begin{itemize}
    \item Group 1: P11 (Computer Science), P06 (Design, Tech, Engineering), P03 (History), P07 (Mathematics)
    \item Group 2: P05 (Design, Tech, Engineering), P02 (Computer Science), P04 (Mathematics)
    \item Group 3: P01 (English), P08 (Computer Science), P10 (Design, Tech, Engineering), P09 (History)
\end{itemize}

Workshop participants engaged in hands-on tasks below to foster creative thinking and group collaboration, with discussions throughout to explore their ideas:

\begin{enumerate}
    \item After establishing workshop norms \cite{ostrowski2021long}, participants used \href{https://miro.com/}{Miro boards} for three five-minute rounds of \textbf{open brainstorming}. They filled post-it notes identifying (1) key challenges in PBL, shared (2) past strategies and tools they had used to address those challenges, and (3) envisioned dream tool features. After brainstorming rounds, participants reviewed the post-it notes, labeled promising ideas, and linked them to past experiences and potential concerns. They then grouped these ideas into categories on the Miro boards, which helped reinforce the ideas that emerged from our interviews.

    \item We then gave a \textbf{presentation about GenAI} with examples of the technology’s use cases in K-12 education. We explained how ChatGPT works and encouraged participants to use this knowledge to explore its potential and limitations in classrooms. Teachers also shared their experiences using these tools in their teaching. {Our teachers gravitated towards using LLMs during subsequent stages of the study, even though they may have referred to GenAI during discussions as it was a more commonly recognized term for them.}
    
    \item Finally, participants expanded on the initial problem scoping to ideate where GenAI {(particularly LLMs)} could be incorporated into their selected category from the board. They created \textbf{storyboards} \cite{buxton2010sketching} of concrete scenarios depicting PBL implementation challenges and a concept of how a fictional LLM tool could serve as a solution in their everyday teaching context. These storyboards were then shared and discussed within the group. 
\end{enumerate}
 
Written material from the participants and notetakers, and audio recordings of the small group discussions were collected online for analysis. 

\paragraph{\textbf{Workshop 2}}

We summarized the brainstorming boards and storyboards from Workshop 1 to guide activities for Workshop 2 that developed conceptual design prototypes for PBL GenAI tools. Each group session consisted of three to four teachers, a primary facilitator, and a notetaker (similar to Workshop 1). We tried to reorganize participant groups to encourage a wider range of discussions across the two workshops: 

\begin{itemize}
    \item Group 1: P11 (Computer Science), P05 (Design, Tech, Engineering), P09 (History)
    \item Group 2: P01 (English), P08 (Computer Science), P10 (Design, Tech, Engineering)
    \item Group 3: P03 (History), P06 (Tech Integration: all subjects), P02 (Computer Science), P04 (Mathematics)
\end{itemize}

\begin{figure*}
    \centering
    \includegraphics[width=1\linewidth]{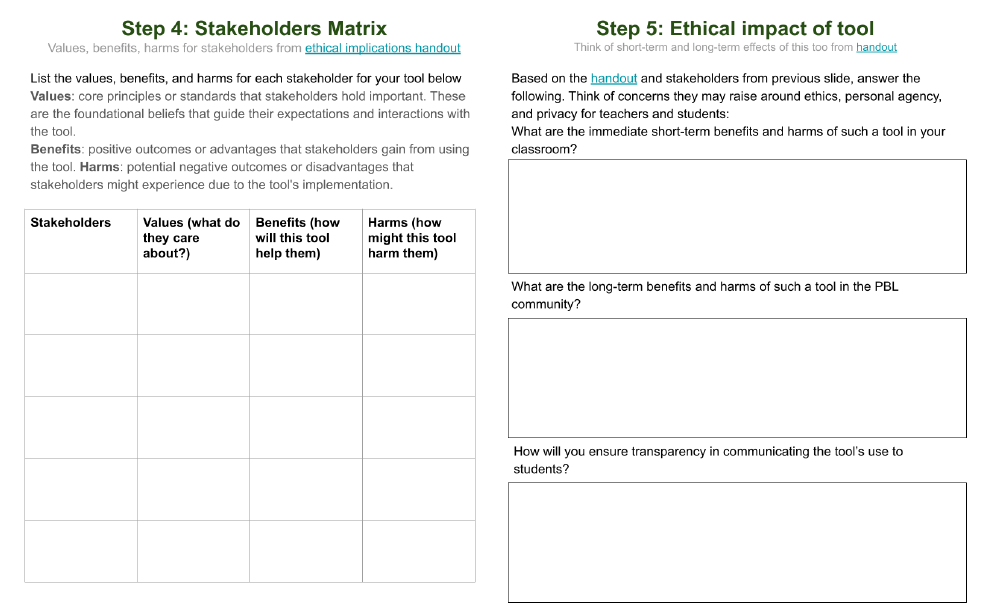}
    \caption{Examples of {blank worksheets distributed to participants in workshop 2} outlining stakeholder impacts and ethical implications of PBL-LLM tools}
    \label{fig:worksheets456}
    \Description{The figure includes two worksheets to guide discussions on developing and integrating a generative AI tool in education, focusing on stakeholders and ethical impacts. Step 4 asks participants to identify values, benefits, and harms for stakeholders. Step 5 explores short- and long-term ethical impacts, including transparency in tool use.}
\end{figure*}

In this workshop, participants developed conceptual prototypes based on the storyboards from Workshop 1. They discussed their tools' LLM roles, its ethical impact on stakeholders, and classroom integration constraints by collaboratively completing worksheets with specific prompts using \href{https://slides.google.com/}{Google Slides}, suitable for the remote setting. An example of {blank} worksheets from some of the steps below is shown in Figure \ref{fig:worksheets456}. We summarize this process below:

\begin{enumerate}
    \item \textbf{Define Final Tool Idea:} Participant groups first selected 1-2 storyboards with overlapping themes for further exploration, wrote a detailed description, and ensured all members agreed on the concept, which became the basis for their final prototype.
    \item \textbf{Identify Stakeholders:} Participants then identified direct and indirect stakeholders involved in their chosen scenarios and their specific roles and interaction with the tool. 
    \item \textbf{Identify How the GenAI Prototype Is Used: } Next, they detailed the role of the GenAI in the selected scenarios. 
    Groups created a list of envisioned features, considered their data-sharing requirements, preferred devices for deployment (phone, tablet, or laptop), and potential integration with existing Learning Management Systems (LMS) like Canvas, Google Classroom, etc.

    \item \textbf{Ethical Implications: }We then led participants through discussions of their designs' societal impacts. Participants received an ethical implications handout to identify values, benefits, and potential harms for each stakeholder group from Step 2 \cite{ali2023ai, arnold2019factsheets, liao2020questioning}. They also investigated the immediate short-term and long-term consequences of using this tool in their classrooms and the PBL community. 
    
    \item \textbf{Classroom Constraints:} Finally, participants considered practical aspects of integrating their prototype in their schools, including resource requirements, tool training and maintenance, anticipated challenges (technical, logistical, pedagogical), and strategies to address these challenges effectively.
\end{enumerate}

Prototype worksheets, facilitator notes, and audio recordings of discussions were collected for analysis. 

\paragraph{\textbf{PBL GenAI Wireframes Walkthrough}}
\label{testing}

Using data from interviews and co-design sessions, we created wireframes outlining valuable design features for teachers using PBL, focusing on curriculum support (project brainstorming and lesson planning), assessment support (rubric creation with differentiation and grading), and progress tracking (monitoring progress at the individual student, group, and class levels). 
These wireframes and support areas, described further in our results section, were reviewed in one-on-one sessions with teachers, who assessed the usability of the tool's wireframes within their classroom contexts and unique pedagogical practices.
Sessions were conducted remotely via Zoom and recorded with consent. We expanded on this feedback by including other teachers outside Study 1 (Section \ref{study2methods}).
We administered a post-survey at the end to understand participants' perceptions of GenAI and PBL and gather their reflections on Study 1's activities.

\subsubsection{\textbf{Data Analysis}}

We transcribed all interviews and took an inductive, iterative approach to thematic analysis \cite{braun2019reflecting}. Initial codes were identified from the transcripts and refined into broader themes, which were continuously reviewed to accurately capture participants' insights. Two researchers then coded the transcripts, achieving a Cohen’s kappa of 0.906 on 40\% of the data, indicating substantial interrater reliability \cite{stemler2019comparison, mchugh2012interrater}. We then deductively applied these themes to the rest of our co-design and wireframe testing data \cite{azungah2018qualitative}. \\

\begin{table*}[ht]
    \centering
    {
    \begin{tabular}{|l|p{0.5\linewidth}|}
    \hline 
         \textbf{Pre-survey metric}& \textbf{Collated particpant responses}\\
         \hline \hline 
         Participants' gender&14 female and 16 male\\\hline
         Subjects taught (multi-select)& Science (15), Mathematics (7), Computer Science (6), Engineering and Technology (4), STEAM (3), English (1), Business (1), and Teacher Education (1)\\ \hline 
         Students' age groups (multi-select)& Kindergarten (2), 1st-5th grade (7), 6th-8th grade (14), 9th grade and above (21)\\ \hline 
         Experience with PBL& 46\% for more than 5 years, 23\% for 3-5 years, and 31\% for a year or less.\\ \hline 
         Comfort levels with PBL (scale 1-5)& 5: 40\%, 4: 26.7\%, 3: 16.7\%, 2: 10\%, 1: 6.7\%\\ \hline
 Describe past PBL experience&Difficult (Significant challenges, poor engagement, and outcomes): 23\% \newline
Neutral (mix of challenges and successes): 46\% \newline
Very successful (Highly positive, engaged students, and excellent outcomes): 33\%\\\hline
 Experience with GenAI tools like ChatGPT&1: 20\%, 2: 26.7\%, 3: 26.7\%, 4: 20\%, 5: 6.7\%\\\hline 
    \end{tabular}
    }
    \caption{Study 2 participants' details, N=30}
    \label{tab:study2_participants}
    \Description{The table provides details of the participants (N=30) from Study 2, which includes teachers with varying levels of expertise in project-based learning (PBL). The table presents collated participant responses on gender, subjects taught, student age groups, experience with PBL, comfort with PBL, description of past PBL experiences and experience with GenAI tools like ChatGPT. There is an equal mix of male (M) and female (F) teachers across different subjects such as Science, Math, Computer Science, Technology, and Engineering, STEAM, English, Business, and Teacher Education. Most participants teach middle and high school. There is a mix on experience and comfort levels. Participants are equally divided between those who had found success with PBL and those who had not. Experience with GenAI tools also varies all across.}
\end{table*}

\subsection{STUDY 2}
{The bulk of the co-design process was conducted with the expert teachers in Study 1. Following this, we conducted a second study involving teachers with varying levels of experience and comfort with PBL (and hence varying expertise as per our definition in 3.1.1). This workshop aimed to gather a wider range of perspectives from a diverse group of teachers, ensuring the tool wireframes could adapt to different teaching styles and classroom environments (as per RO2).}
This approach also evaluated the tool's potential to reduce barriers {of entry} to PBL and allow educators to focus more on experiential learning and holistic development.



\subsubsection{\textbf{Participants and Recruitment}}
\label{study2methods}

Thirty STEM teachers from diverse U.S. states and school contexts were recruited in collaboration with the MIT Scheller Teacher Education Program and MIT Alumni Clubs. These teachers, admitted to the Science and Engineering Program for Teachers (SEPT) program, participated in a 2-hour workshop on GenAI for project-based and collaborative learning. 
We distributed a pre-survey to explore the backgrounds, {subjects,} and experiences of workshop participants {(see Table \ref{tab:study2_participants}). Teachers primarily taught middle and high school students and had different levels of PBL experience (average of 3.4 years) and comfort (average of 3.8/5) resulting in varying expertise.}
Participants were equally divided between those who had found success with PBL and those who had not.



\subsubsection{\textbf{Data Collection}}

Study 2's workshop started with brief discussions on participants' perceptions of GenAI and collaborative learning. Following these, participants were divided into three groups based on their pedagogical interests: curriculum planning, assessment, and progress monitoring.
Similar to our procedure at the end of section \ref{testing} of Study 1, we walked Study 2 participants in our small group sessions through the same set of wireframes, which focused on curriculum support, assessment support, and progress tracking. After a brief review, discussions centered on the specific support area participants in the small group had selected. Participants explored the wireframes, provided feedback, and assessed usability in their classroom contexts. They also identified tool tasks best kept human-driven, and their impact on teaching creativity, well-being, workload, and stress.
We followed this with a post-survey to gauge participants’ excitement and concerns on GenAI, interest in PBL, desired additional features for the wireframes, and onboarding resources needed for classroom use.



\subsubsection{\textbf{Data Analysis}}

Discussions from all three groups were audio recorded with consent from participants for the purposes of transcription. We then deductively applied the same themes from Study 1 to code our data from Study 2 \cite{azungah2018qualitative}. \textbf{For the purposes of this paper, we focus on Study 1's interviews and co-design workshops, as well as the PBL GenAI wireframes and their feedback from Study 1 and Study 2.
Note that the feedback on the wireframes has been analyzed and discussed by combining the results from both studies, as they together shaped the design guidelines and future directions for the tool.}

\begin{table*}[ht]
    \centering
    \begin{tabular}{|p{0.2\linewidth}||p{0.2\linewidth}|p{0.5\linewidth}|}\hline
  {\textbf{Category}}&\textbf{Themes} & \textbf{Definitions}\\\hline \hline 
  {\multirow{15}{*}{(A) Set stage for PBL needs }}& school context& details on subjects taught, grades, public/private school, student backgrounds, teacher experience, the way PBL is structured in the school\\ \cline{2-3}
          & specific project examples& specific examples that teachers gave on projects they have implemented in their classrooms. Include details about the project sp we know what is, not just in reference to other parts pedagogy.\\ 
          \cline{2-3}& student agency& how teachers give students autonomy, choice, freedom on projects, including group formation\\ 
                    \cline{2-3}& student differentiation & how material/grading/instruction is adapted for different groups/needs of students, including personalized support. Supports for some students.\\ 
                    \cline{2-3}&student scaffolding& any and all tools/strategies for scaffolding different parts of PBL for students. Supports all students get.\\  
                    \cline{2-3}&teacher role& role of teacher in PBL setting\\ 
                    \cline{2-3}&personalized feedback& how individual and personal feedback is given to every student\\ 
                    \cline{2-3}&goals, checklists, progress tracking& anything around setting goals, milestones, checklists, deadlines, monitoring regular progress etc. (exit tickets)\\  
                    \cline{2-3}&online learning platforms and tools& types of online learning platforms and grading platforms used (e.g. Canvas, Google Classroom)\\  
                    \cline{2-3}&rubrics/grading scales& details on rubrics and grading schemes/strategies used for evaluating different parts of projects \\ 
            \cline{2-3}&self and peer reflections&any details on self and peer reflections/feedback as part of projects\\
             \cline{2-3}&managing student groups&details on managing student groups and dynamics, interpersonal skills and interactions. conflict resolution, assigned roles, leveraging different strengths etc.\\ 
            \cline{2-3}& remote PBL&experiences and challenges of PBL in online settings. Includes structure and procedures for conducting class\\
           \cline{2-3}& PBL inclusion&details on gender gaps, STEM, equity, etc.\\ 
           \hline \hline
{\multirow{4}{*}{(B) Challenging aspects}}&PBL challenges for students&specific details on challenges encountered by students in PBL\\ \cline{2-3}
  &PBL challenges for teachers&specific details on PBL challenges encountered by teachers\\ \cline{2-3} 
  &PBL teacher perceptions&teacher perceptions of PBL\\ \cline{2-3}
  &PBL student perceptions&student perceptions of PBL\\ \hline \hline
{\multirow{3}{*}{(C) AI + PBL }}&AI perceptions&any details on teacher's perceptions/details of AI, including GenAI \\ \cline{2-3} 
  &current AI uses/challenges in classroom&details on use cases and challenges from using AI in classrooms\\ \cline{2-3} 
  &PBL wish list&details on things teachers wish for in the future in their classrooms\\ \hline 
    \end{tabular}
    \caption{Final set of themes and their definitions from our data analysis. {These span three main types: (A) those that set the stage for PBL needs, (B) those that describe challenging aspects of PBL implementation, and (C) those that show AI integration in PBL}}
    \label{tab:themes}
    \Description{The table outlines key themes and their definitions related to project-based learning (PBL) and AI in education. It includes themes like school context, student agency, teacher roles, and challenges faced by both students and teachers in PBL. It also addresses the use of online learning platforms, AI perceptions, and inclusion issues such as gender gaps and equity in PBL settings. Themes have been classified by category: (A) those that set the stage for PBL needs, (B) those that describe challenging aspects of PBL implementation, and (C) those that show AI integration in PBL.}
\end{table*}

\section{RESULTS}

\begin{figure*}[ht]
    \centering  \includegraphics[width=0.9\linewidth]{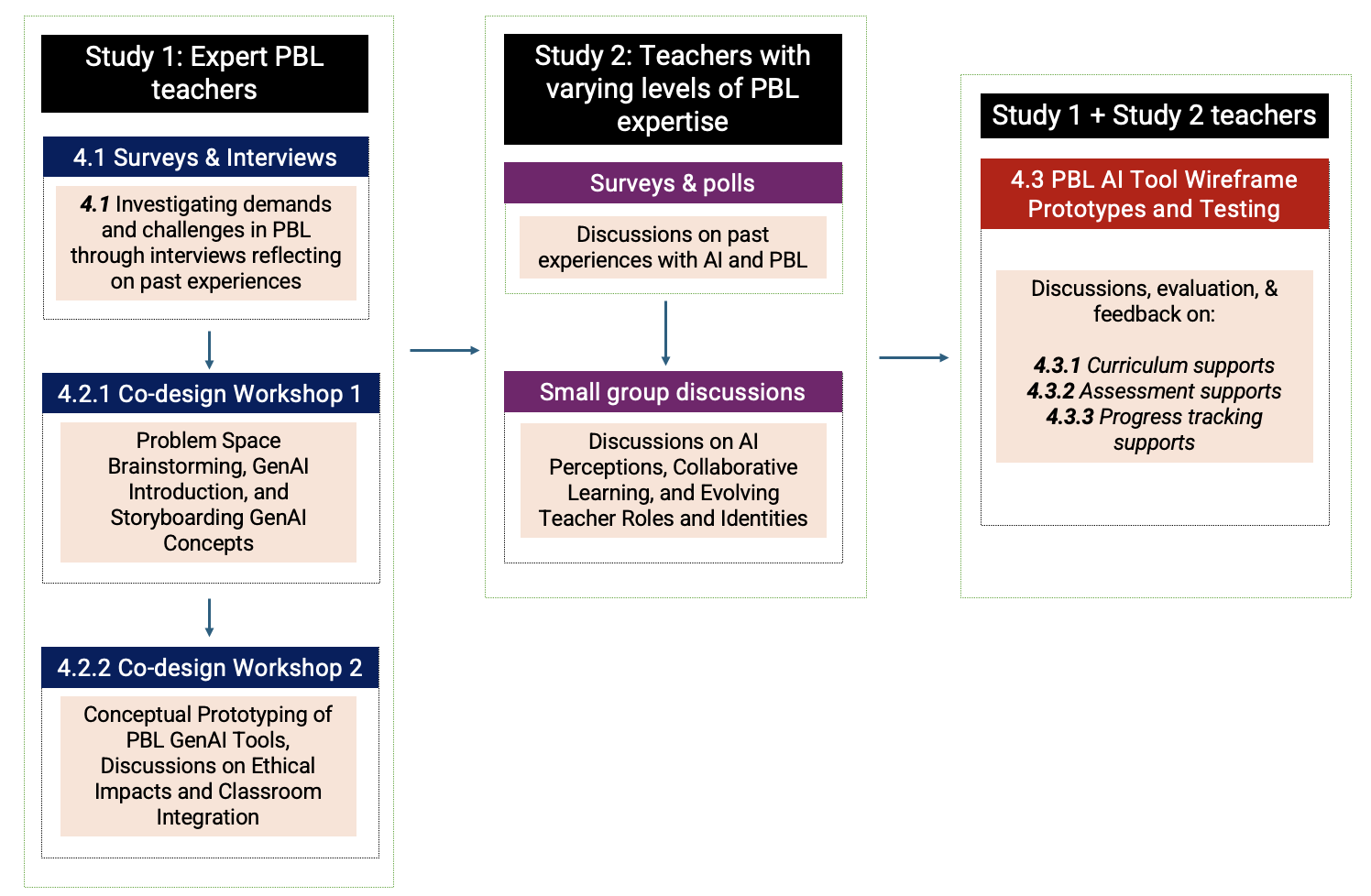}
    \caption{A timeline illustrating the various stages of our study alongside the corresponding paper section numbers where their findings are discussed. For Study 2, note that we present findings from the final wireframe testing phase only, as mentioned in Section 3.2.3}
    \label{fig:methods-overview}
    \Description{The figure provides a timeline of two studies in our paper. Study 1 involves expert PBL teachers and begins with surveys and interviews to gather insights into their demands and challenges in implementing PBL. This is followed by two co-design workshops: the first workshop centers around problem space brainstorming, an introduction to generative AI (GenAI), and the storyboarding of GenAI concepts, while the second workshop focuses on conceptual prototyping of PBL GenAI tools, with discussions on ethical impacts and classroom integration. Study 2 engages PBL teachers with varying levels of expertise through workshop surveys and polls to collect their experiences with AI and PBL. Additionally, small group discussions explore their perceptions of AI in collaborative learning and how it influences their roles and identities as teachers. The findings from both studies are integrated into the iterative development and testing of wireframe prototypes for a new LLM tool designed to support PBL. Teacher feedback from both study participants is used to refine the tool's design around curriculum, assessments, and progress tracking supports.}
\end{figure*}
\begin{figure*}[ht]
    \centering    \includegraphics[width=0.6\linewidth]{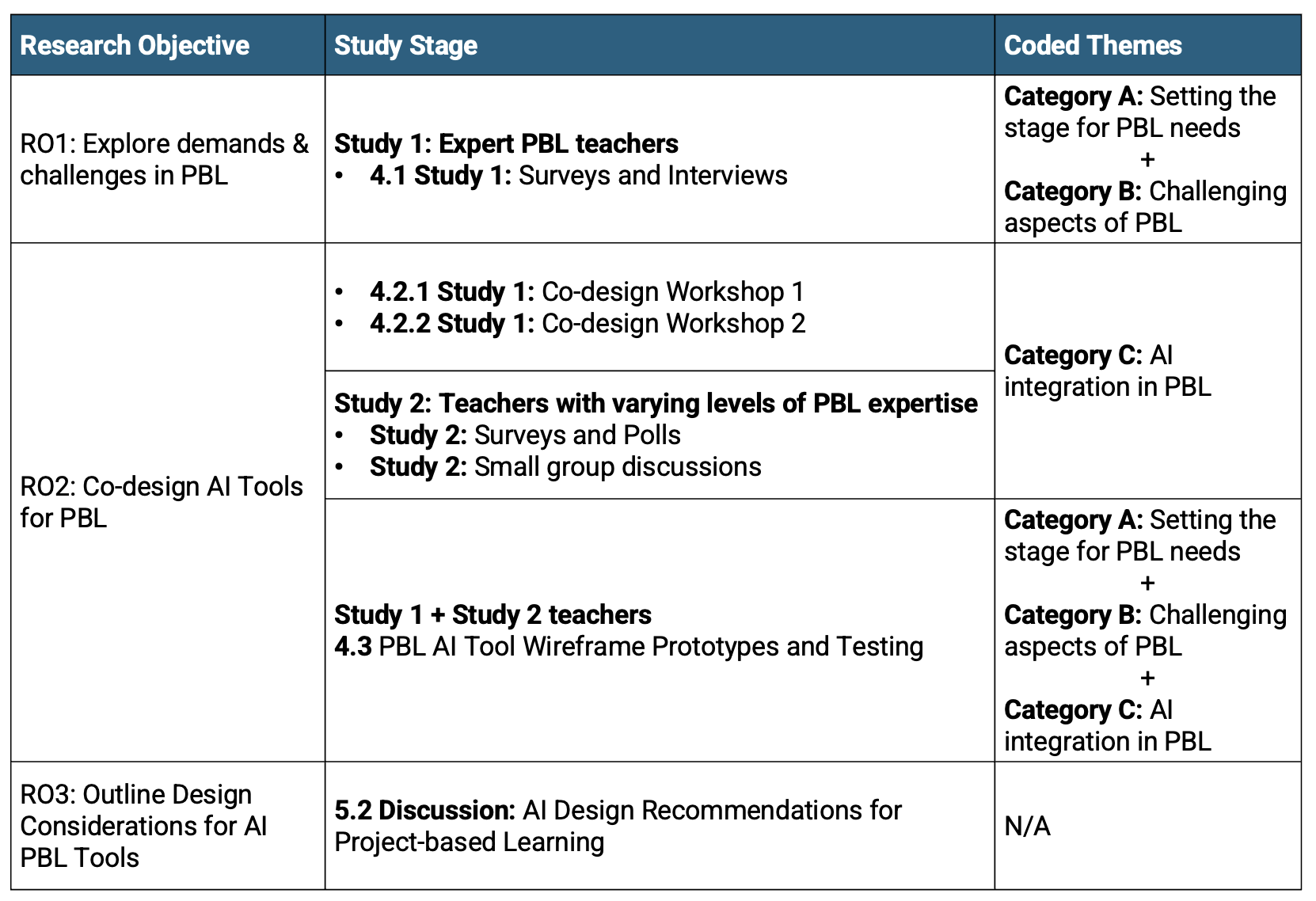}
    \caption{A figure displaying the mapping of research objectives to the different stages of the study and their results' coded themes}
    \label{fig:mapping}
    \Description{This figure displays the mapping of research objectives to the different stages of the study and their results' coded themes. RO1 is explored through study 1 surveys and interviews with expert teachers. These give rise to themes in categories A and B. RO2 is explored through study 1’s co-design workshops with expert teachers, study 2’s surveys, polls and small discussions with teachers of varying levels of PBL expertise. This leads to the emergence of category C themes. Additionally, RO2 is also examined using the PBL-LLM wireframes evaluated with both study 1 and study 2 teachers. These wireframes see themes from all three categories A, B and C. Finally, RO3 is explored in the design recommendations subsection of the paper under Discussion.}
\end{figure*}

Table \ref{tab:themes} shows themes and their corresponding definitions that emerged from our data. {Section 4.1 delves into Category A themes that set the stage for foundational PBL needs and Category B themes regarding challenging aspects associated with addressing these needs. Section 4.2 highlights Category C themes that suggest pathways for integrating GenAI (specifically LLMs) into PBL, drawing on key insights from our two co-design workshops. Finally, Section 4.3 presents wireframes for our proposed PBL LLM tools, synthesizing themes from all three categories. This mapping of research objectives, study stages, and themes is shown in Figures \ref{fig:methods-overview} and \ref{fig:mapping}.}

\subsection{Exploring Demands and Challenges in Project-Based Learning {(Category A and B themes)}}
We present findings from our interviews in Study 1, focusing on themes highlighting teachers' core beliefs about the demands of high-quality PBL implementation {(Category A in 4.1.1)} and the challenges for them and their students {(Category B in 4.1.2). These answer RO1}. { We recognize that some of the findings from these interviews align with themes already discussed in the literature. However, our primary aim was to build relationships with our teachers by gaining a nuanced, in-depth understanding of the types of projects they implement within their distinct subject areas, along with the specific challenges they face. By centering teachers' voices and lived experiences, we capture and communicate the authentic context of their pedagogy, enabling readers to better appreciate how the subsequent co-design artifacts are shaped by and respond to their unique classroom demands.}

\subsubsection{\textbf{Demands in Project-Based Learning}}
Teachers in our study emphasized the distinct elements of PBL that had to be addressed to foster engagement and promote deeper learning.
P10 stressed the importance of focusing on the big picture while managing granular project details, helping students see the connections between project components.
Teachers underscored the interdisciplinary nature of PBL, requiring them to develop the expertise to integrate subjects like physics, engineering, and design, to foster holistic learning. 
The \textbf{adaptability and improvisational aspects of PBL} were recurrent themes, as illustrated by P06:

\begin{quote}
    \textit{“Last year, I was ready to throw my 3D printer out the window because students couldn't even name their projects properly despite detailed instructions. I had to rethink my approach and make tweaks...Flexibility and recognizing that you don't have everything figured out is key to successful PBL.” -- P06} 
\end{quote}

Teachers shared how \textbf{prior PD experiences}, including workshops and training, were crucial for implementing PBL effectively and staying current with evolving educational practices and technologies.
They also spoke about their roles and responsibilities as \textit{\textbf{``expert facilitators and coaches''}}, helping students navigate projects while promoting independence.
P02 noted that much of the time was spent observing, asking questions, and offering support as needed, allowing students to explore and experiment. Handling multiple projects required effective \textbf{context-switching and organization.} P07 stressed building student confidence, particularly in challenging subjects like math, and creating a supportive classroom where mistakes were part of learning. 
Our teachers relied on standardized rubrics and project descriptions to ensure fairness, but they had to \textbf{adapt the content to different abilities}, necessitating the creation of multiple versions of resources.
Teachers also understood that some students preferred hands-off approaches while others needed frequent support, and they balanced these needs through annotated monitoring and \textbf{personalized interactions}:

\begin{quote}
\textit{"Some students embrace it and love the challenge and freedom to explore in their own way. Others resist and rebel against it, preferring a more traditional approach with worksheets, lectures, and tests. And then there's everything in between. Some are willing but don't know how to do it." --P03}
\end{quote}

Teachers used \textbf{scaffolding tools like checklists} to help students adapt to PBL's self-directed approach, keeping them organized and focused. They mapped rubric items to content standards and other criteria, such as creativity and sustainability, refining these annually to match evolving goals. To manage group projects, teachers proactively defined tasks and deadlines for individual members, conducted \textbf{regular check-ins}, and \textbf{encouraged reflection}, ensuring equitable contributions and progress. These strategies fostered a structured yet flexible learning environment, enhancing students' time management, self-assessment, problem-solving, and metacognitive skills.

\subsubsection{\textbf{Challenges in Project-Based Learning}}
Implementing PBL presented organizational difficulties for teachers, balancing content depth and breadth, and managing diverse student needs.
The shift from traditional assessments to more dynamic, project-based evaluations also demanded innovative assessment methods. P02 highlighted the challenges of \textbf{evaluating open-ended projects without additional support} staff in schools. 
Teachers grappled with balancing the depth of knowledge required for specific projects with the breadth of content mandated by the curriculum. This often meant that students developed an in-depth understanding of their chosen project topic but lacked comprehensive knowledge of other areas, impacting their performance on traditional assessments.  
Integrating technology also added complexity, with students' enthusiasm for new tools sometimes \textbf{diverting focus from project goals}. Additionally, the availability of resources and the need to create or adapt materials presented ongoing struggles. Teachers often felt constrained by their curriculum's scope and sequence, which limited their ability to fully embrace PBL's potential.
\textbf{Fair grading} was particularly challenging due to the subjective nature of projects and the need for continuous feedback. P10 highlighted the \textbf{workload} involved:

\begin{quote}
\textit{"It's a lot of work. I'm constantly grading. I don't get a break because it's a continuous cycle of work. I give iterative feedback throughout, and my turnaround is as quick as possible, but it means grading all weekend." -- P10}
\end{quote}

\textbf{Maintaining student engagement} was another recurring challenge in PBL. P01 noted that some students (e.g. struggling readers) found it hard to stay on task, requiring additional prompts and support to remain focused. 
\textbf{The transition from lecture-based learning to PBL} was difficult, with students initially resisting the shift from teacher-directed instruction. P05 recounted how some students found projects daunting and needed the \textbf{tasks broken into smaller, manageable steps}. This step-by-step approach helped students gradually build confidence and see their progress, ultimately leading to successful project completion. Teachers also stressed the importance of social-emotional development, noting the pandemic's negative impact on students' ability to collaborate and support each other.  Flexibility and empathy were essential in helping students manage PBL challenges, particularly when overwhelmed or facing mental health issues. P09 observed that students often struggled to apply memorized concepts to projects, requiring a structured approach with front-loaded skills before diving into projects. The hands-on nature of PBL demanded continuous support to bridge the gap between conceptual knowledge and practice. P11 noted challenges with \textbf{unequal team participation}, affecting grades and contributions. Ensuring fairness required careful monitoring and regular teacher intervention. 

These challenges in student engagement compounded in \textbf{remote settings}, with many students preferring chat over cameras or microphones. P05 used breakout rooms for small group discussions and maintained engagement through Zoom polls and exit tickets, despite inconsistent Wi-Fi and varying school district permissions. \textbf{Logistical barriers, like inconsistent access to resources}, further complicated remote PBL.

\subsection{Integrating AI in Project-Based
Learning {(Category C themes)}}

Building on interviews, we present findings from the co-design process, where teachers explored customizing LLM tools to meet PBL's specific demands {(Category C). These help answer RO2.}


\subsubsection{\textbf{Co-design Workshop 1: }}
The first workshop aimed to identify challenges teachers faced with implementing PBL and map scenarios in which LLMs tools could help. 

\paragraph{\textbf{Brainstorming- Challenges, Current Strategies, and “Magic wand” features:}} Across our three groups and rounds of brainstorming, we saw similarities in the categories that emerged. 

Participants discussed strategies for \textbf{setting goals and tracking progress} in PBL. One suggestion was a \textit{"dashboard to monitor progress with a traffic light system (red, yellow, green)" (Group 1)}, for visual tracking. Another idea involved using a checklist in the form of a \textit{"bingo board for final projects" (Group 3)}. However, these tools sometimes failed when items were checked off without actual completion. Frequent check-ins and breaking projects into manageable tasks were also mentioned, though this process was noted to be labor-intensive due to the need for continuous tracking, timely feedback, and adjustments to lesson plans.
Participants promoted \textbf{student autonomy} by \textit{"maintaining momentum” (Group 3)} during classroom time, \textit{“keeping project ideas fresh" (Group 1)}, and organizing \textit{“fun, creative tasks within projects to reduce cognitive and social discomfort” (Group 1)}. Allowing students to \textit{"pick goals with teacher feedback" (Group 1)}, and “\textit{outline the top three criteria they wish to be graded on” (Group 3)} gave them more control over their learning. However, balancing this autonomy with meeting curriculum standards remained a challenge.

Participants emphasized the need for \textbf{differentiation and personalization} by \textit{"challenging all students where they are" (Group 1)} and providing \textit{"feedback targeted at their current level of maturity, technical skill, and aspirations" (Group 1)}. There was a call for tools that offer \textit{"additional resources tailored to students' strengths and weaknesses" (Group 3)} and help navigating complex materials. Ensuring \textit{"equitable [opportunities] for students of diverse backgrounds and learning levels." (Group 3)}, along with having alternate plans if needed, was seen as crucial for student success.

\textbf{Managing group dynamics} was another area of focus within the brainstorming boards, such as dealing with \textit{"students who give minimal effort to the team" (Group 1)} and \textit{"ensuring that everyone in the group understands the tasks" (Group 2).} \textit{"Assessing [group] projects for the group as well as individuals" (Group 2)} was noted as a difficulty, highlighting the need for fair rubrics. \textbf{Time management} also surfaced as a significant concern. While self-reflecting on incremental progress was seen as valuable for students, it was also time-consuming. Integrating technology required time to train students, and providing meaningful feedback posed a challenge when balancing thoroughness with efficiency. 

Participants discussed various \textbf{grading strategies and challenges}, such as the use of external automated tools for scaffolded templates like \textit{"Repl.it, Codecheck.it, CodingRooms, Google Colab Notebooks" (Group 2)}, which were helpful but time-consuming to set up. There was debate over balancing and \textit{"assessing [final] product vs. [learning] process"} \textit{(Group 2)} within projects. 
Other concerns included integrating standards-based grading into rubrics while managing the pressure from parents to convert these into traditional letter grades for transcripts. Fairness in grading was also highlighted, given the complexity of creating rubrics for different project parts, student needs, and interdisciplinary topics. Some educators suggested involving students in the evaluation process through \textit{"self-grading along with teacher grading" (Group 3)}, which promoted student self-awareness and provided deeper insights to educators for personalized feedback. 

\paragraph{\textbf{Storyboarding Areas of Opportunity for GenAI Tools in PBL:}} Participants expanded on initial problem scoping by creating storyboards illustrating scenarios in which a hypothetical LLM tool could help address challenges in their daily teaching practices. Our analysis revealed three major areas of support: \#1: curriculum and lesson planning, \#2: assessments and grading, and \#3: managing group dynamics and progress tracking. \\ 

\textbf{[Support area \#1]- Curriculum \& Lesson Planning + [Support area \#2]- Assessments \& Grading}:
Our participants created storyboards where teachers and students used the LLM system to collaboratively brainstorm project ideas, generate lesson materials, and customize these for diverse student needs. They also envisioned the LLM aiding in creating assessments (rubrics), and automating grading with teacher input. The themes from support areas \#1 and \#2 frequently co-occurred in these scenarios. For example: 

\begin{figure*}
    \centering
    \includegraphics[width=1\linewidth]{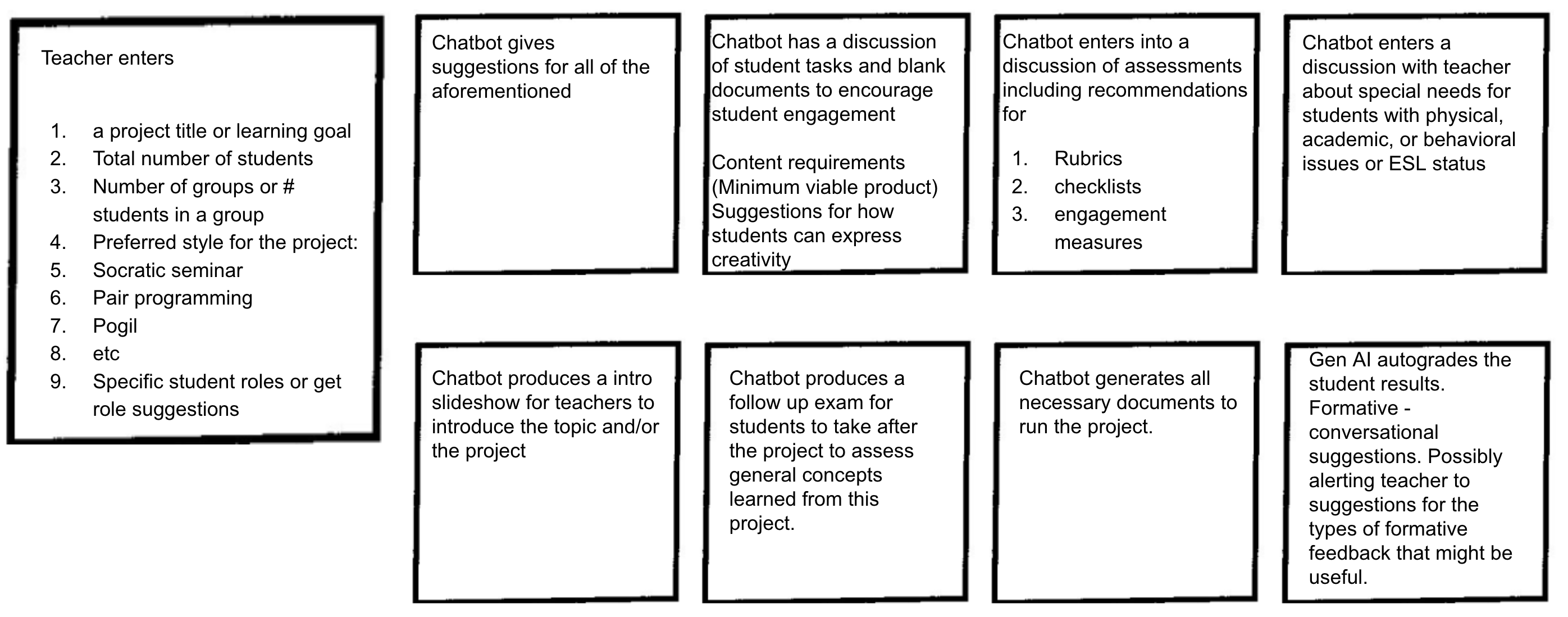}
    \caption{Storyboard (P11) mapping a step-by-step teacher-LLM interaction for project and assessment ideation}
    \label{fig:storyboard-P11}
    \Description{The figure illustrates a storyboard by P11 outlining the process of an LLM tool assisting teachers in planning and implementing a PBL activity. It shows steps where the teacher inputs project details, and the LLM provides suggestions for project design, student tasks, assessments, and accommodations for special needs. The LLM generates various resources, such as an introductory slideshow, exam, and necessary documents, and also autogrades student results, offering formative feedback and recommendations.}
\end{figure*}
\begin{figure*}
    \centering
    \includegraphics[width=1\linewidth]{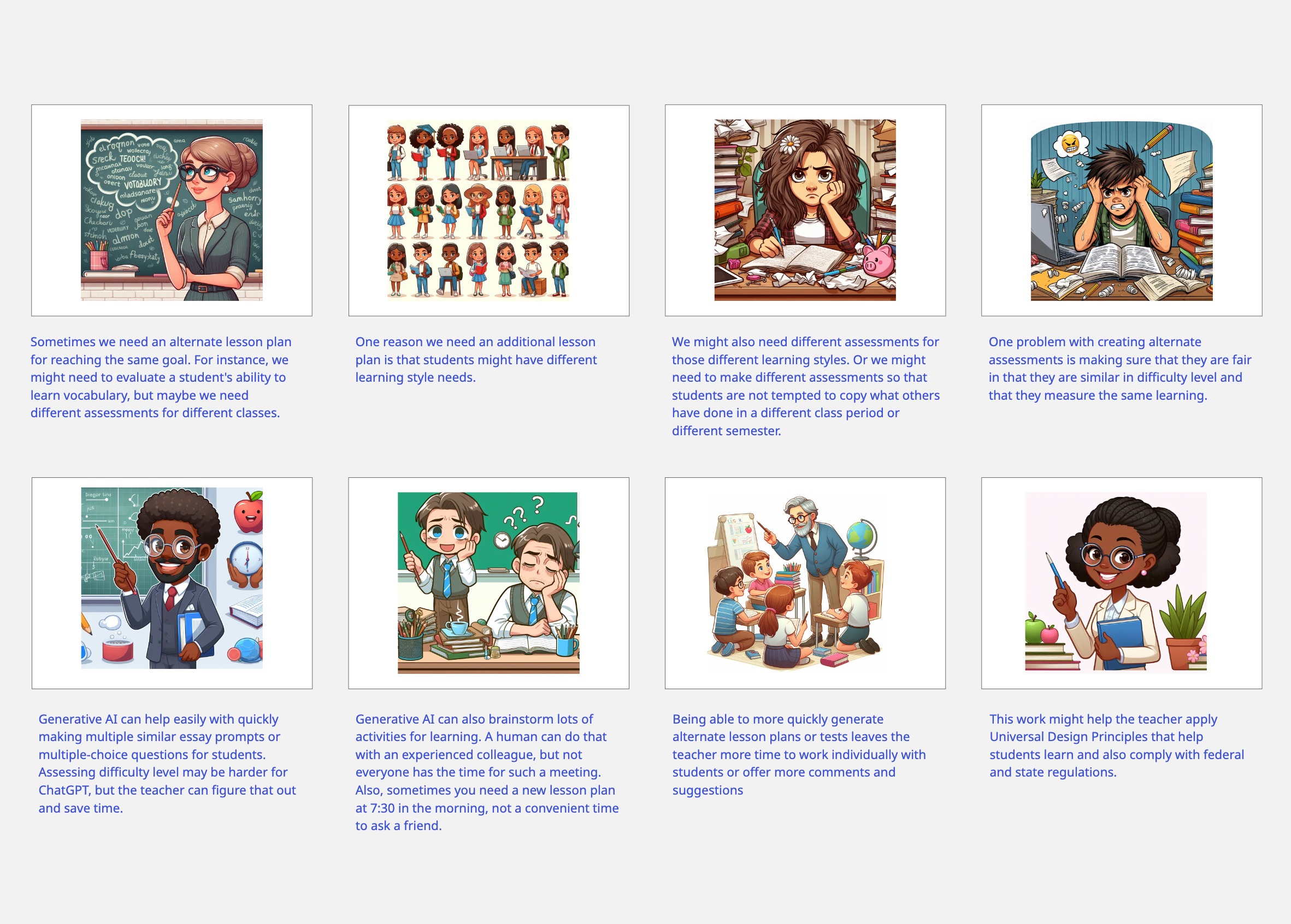}
    \caption{Storyboard (P05) showing an LLM tool for alternative assignments and lesson plans}
    \label{fig:storyboard-P05}
    \Description{The figure showcases a storyboard from P05 with LLMs for creating alternative lesson plans and assessments. It highlights challenges, such as accommodating different learning styles and maintaining fairness in assessments, and suggests that LLMs can help quickly generate multiple prompts, brainstorm activities, and save teachers time. It also emphasizes that LLM tools can support teachers in applying Universal Design Principles and meeting regulatory requirements.}
\end{figure*}

P11 highlighted the challenges of defining projects and designing assessments in PBL, which can be time-consuming (upwards of 40 hours). They noted difficulty in finding comprehensive resources online suited to various instructional needs. To address these issues, P11 envisioned a LLM-powered chatbot that assists in \textbf{project creation} by generating tailored project plans based on class information and goals provided by the teacher (Figure \ref{fig:storyboard-P11}). The tool would offer suggestions for student tasks, sentence starters, and \textbf{help design rubrics and checklists}, accommodating \textbf{differentiated instruction} by considering students' physical, academic, and behavioral needs. Additionally, it could generate documents like timelines, visual progress indicators, and follow-up exams. Other teachers suggested integrating an open-source database with both tested and untested unit plans for cross-referencing into this design. This storyboard streamlines labor-intensive tasks in PBL, allowing teachers to adapt projects instead of starting from scratch.

\begin{quote}
\textit{“For formative assessments, we need status checks to ensure students are on track during a project. For example, in a 3-day project, students might stay focused because the deadline is near. But in a longer project, they might get off track. The chatbot could provide feedback on their progress, motivation, and any issues they're facing, helping them stay on course... For summative assessments, I would want the chatbot to suggest feedback and explain its reasoning, allowing me to modify it as needed. Feedback should be age-appropriate and level-appropriate. For instance, some students need positive feedback, while others might be motivated by challenges. The chatbot should offer tailored suggestions and allow me to adjust them before finalizing.” -- P11}
\end{quote}


P05’s storyboard (Figure \ref{fig:storyboard-P05}) envisioned LLMs helping \textbf{create} \textbf{alternative or customized assessments} to accommodate different learning styles. They highlighted the challenge of creating varied assessment formats, such as multiple versions of vocabulary tests, to meet diverse needs and prevent sharing of test content between different class periods. P05 suggested LLMs could quickly generate similar essay prompts or multiple-choice questions, \textbf{maintaining fairness} across classes and compiling these into a centralized question bank. They also saw LLMs as a \textbf{brainstorming partner for generating activity ideas} when time is limited, allowing teachers to focus more on individualized instruction and feedback. However, they noted concerns about accessibility, particularly in online settings (pertinent to P05) , where account restrictions limit access to GenAI tools like Google’s Gemini or Microsoft’s Copilot. Other participants discussed the learning curve in using GenAI tools like ChatGPT, noting the difficulty of recalling effective prompts and the need for prompt-sharing among educators for P05's tool. \\

\textbf{[Support area \#3]- Managing Group Dynamics and Progress Tracking}:
We observed storyboards illustrating LLM systems to monitor individual and group progress, streamline communication with students and parents, document project artifacts, track individual contributions within groups, ensure equitable participation, and resolve group conflicts.

\begin{figure}
    \centering
    \includegraphics[width=1\linewidth]{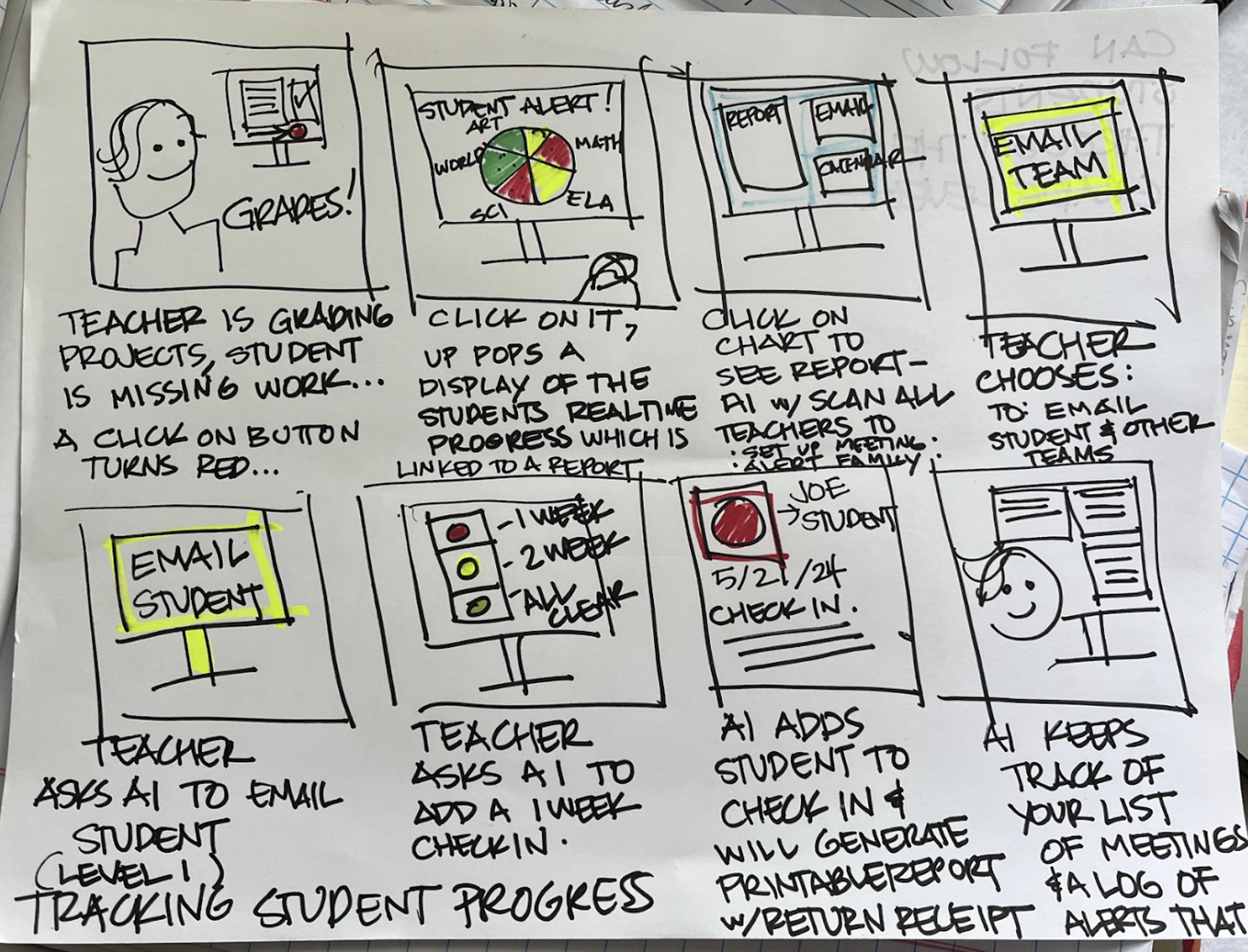}
    \caption{Storyboard (P10) depicting an LLM assistant to monitor at-risk students}
    \label{fig:storyboard-P10}
    \Description{The figure is a storyboard by P10 depicting a process for tracking student progress using LLMs. It shows a teacher grading projects and noticing a missing assignment, prompting a red alert button. The teacher clicks to view a student's real-time progress and options to email the student or other team members. The teacher uses LLMs to schedule a check-in, track meetings, send notifications, and maintain a log of actions, ensuring consistent communication.}
\end{figure}

For example, P10 envisioned an LLM system to \textbf{support teachers monitoring at-risk students} (Figure \ref{fig:storyboard-P10}) by offering a comprehensive visual representation of a student’s academic performance, highlighting key issues like missed assignments. The system would generate detailed reports and \textbf{automate communication with colleagues, parents, and administrators}. The LLM could also help schedule meetings by coordinating calendars and setting reminders for follow-ups to facilitate timely interventions. 
P10 emphasized the value of LLMs in maintaining continuity of support as students progress through different grade levels, analyzing patterns such as chronic lateness or incomplete projects to enable proactive support. Other participants suggested that the tool could also be adapted for student leaders to \textbf{track team submissions }and manage project delays, to avoid adding to the teacher's workload.

\subsubsection{\textbf{Co-design Workshop 2: }}

In this workshop, participants built a conceptual prototype based on storyboard ideas from Workshop 1. 
These prototypes aligned with the three support areas from the storyboards above. We present the discussions on tool stakeholders, LLM data sharing, classroom integration, and ethical impacts below.

\paragraph{\textbf{Tool Stakeholders:}}
Participants identified key groups involved in implementing LLM tools in PBL, outlining their values, potential benefits and harms: \textbf{Classroom teachers, co-teachers, }and\textbf{ new/pre-service teachers} could benefit from reduced administrative workload and support for lesson planning but would be concerned about over-reliance on LLMs and loss of control. \textbf{Special education teachers} emphasize meeting diverse student needs and Individualized Education Program (IEP) goals, with worries about data privacy and maintaining personal connections with students. Participants also discussed involving \textbf{IT teams, the Board of Education, students, }and\textbf{ parents} as stakeholders, each valuing student progress, data security, and educational equity, but wary of privacy concerns, technological divides, and the potential impact on teacher-student interactions.

\paragraph{\textbf{Data sharing requirements of LLM prototypes:}}
Participants suggested that the LLM tool should \textbf{access teacher-specific information}, such as state/local standards, project goals, past assignments, rubrics, guiding questions, and preferred instructional methods (e.g., Socratic seminars, Process Oriented Guided Inquiry Learning (POGIL)). This data would help the LLM align its outputs with educational objectives and teaching styles. To personalize content, educators also stipulated that the system have \textbf{access to comprehensive student profiles}, including standardized test scores, reading and math levels, and accommodations from IEPs. This would enable the LLM to tailor materials to match students' learning levels and accommodations, such as adjusted deadlines for students who require extra time.

Participants also emphasized the importance of \textbf{integrating the LLM tool with existing Learning Management Systems (LMS)} to allow access to student demographic data, schedules, performance records, and adaptation of outputs between teachers. This would also facilitate communication with students and families regarding academic progress and interventions. They acknowledged the need for data to enhance LLM functionality but raised concerns about privacy, especially regarding sensitive data like student learning needs or disabilities. Authorization and \textbf{compliance with data privacy laws} (Ed Law 2d, COPPA, FERPA, HIPAA) was also emphasized.
They also discussed challenges in ensuring the LLM's ability to respect privacy during data access and \textbf{teacher training to use the tool ethically} and responsibly.

\begin{figure}
    \centering
    \includegraphics[width=1\linewidth]{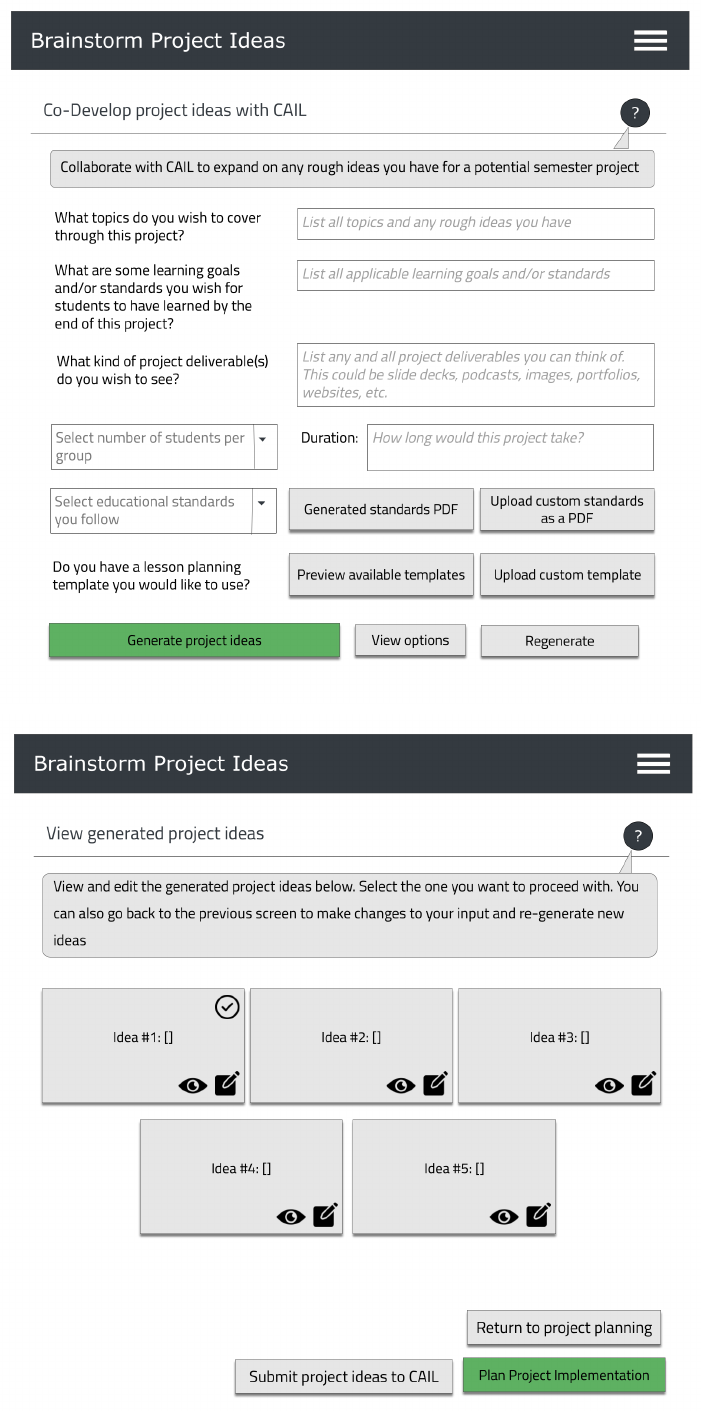}
    \caption{PBL LLM tool supports for brainstorming project ideas}
    \label{fig:project-wireframe}
    \Description{The figure shows a wireframe for a tool to brainstorm project ideas with LLMs. On the left, it allows teachers to input details such as project topics, learning goals, deliverables, group sizes, and educational standards to generate project ideas. On the right, the generated project ideas are displayed for review and editing, with options to select, modify, or regenerate ideas and proceed with planning the project implementation.}
\end{figure}

\paragraph{\textbf{Classroom Integration and Constraints}} Teachers also explored the practical considerations for implementing their prototypes in their specific school and classroom contexts. 
They emphasized needing \textbf{reliable infrastructure}, including strong internet and compatible devices. 
\textbf{High-quality PD and ongoing support} were deemed essential for ethical and practical tool use. A continuous feedback mechanism for user experience and feature requests was also highlighted, along with clear expectations and accountability for its implementation, potentially overseen by tech committees or school boards.

However, teachers anticipated several challenges accompanying this integration, including \textbf{obtaining approval and managing costs}, navigating budget constraints, and ensuring affordability and accessibility. There were concerns about \textbf{resistance to change from educators accustomed to traditional teaching methods}, especially if the tool changes established workflows. \textbf{Time constraints} were also noted, as teachers already face significant demands and may struggle to find time for training. \textbf{Communicating effectively with parents} and addressing their concerns about LLM tools was deemed crucial, with an emphasis on building trust and educating them on benefits and potential harms.

\paragraph{\textbf{Ethical Impact of GenAI Prototypes: Benefits and Harms}}
Participants were enthusiastic about the tool's potential to support teachers by streamlining the creation of projects and rubrics based on educational standards, goals, and \textbf{differentiated practices}, thereby improving lesson planning flexibility and fostering equity in the classroom. Additionally, the tool could enhance student learning by making lessons more interactive, personalized, and connected to real-world problems, potentially \textbf{increasing motivation} and \textbf{participation}. Participants, however, expressed concerns about the tool potentially reducing teacher and student autonomy through \textbf{over-reliance} on LLM-generated recommendations, which could undermine human judgment and diminish valuable teacher-student interactions crucial for trust and understanding. They also worried about the tool \textbf{disrupting existing workflows} during its initial integration and \textbf{exacerbating digital inequalities} due to varying student access to reliable internet.

Participants saw the tool as a way to make \textbf{PBL more approachable} for teachers, encouraging continued use and innovation by simplifying the process. They also noted its potential to maintain consistency across classes and \textbf{align educational materials with state standards}, potentially enhancing outcomes. On the other hand, they were concerned that over-reliance on LLMs could stifle creativity, resulting in generic responses and reducing originality in classroom activities. They also raised issues about potential \textbf{copyright violations}, loss of creative ownership in education, and the invasiveness of the technology, especially in material sourcing.
Participants emphasized the need for transparency in LLM tool use, including clear communication with parents, administrators, and other stakeholders. They stressed understanding the LLM's data sources and algorithms to avoid issues surrounding its \textit{\textbf{"black box"}} nature and suggested incorporating warnings into interfaces to \textbf{verify LLM-generated content} before usage. 

\begin{figure*}
    \centering
    \includegraphics[width=1\linewidth]{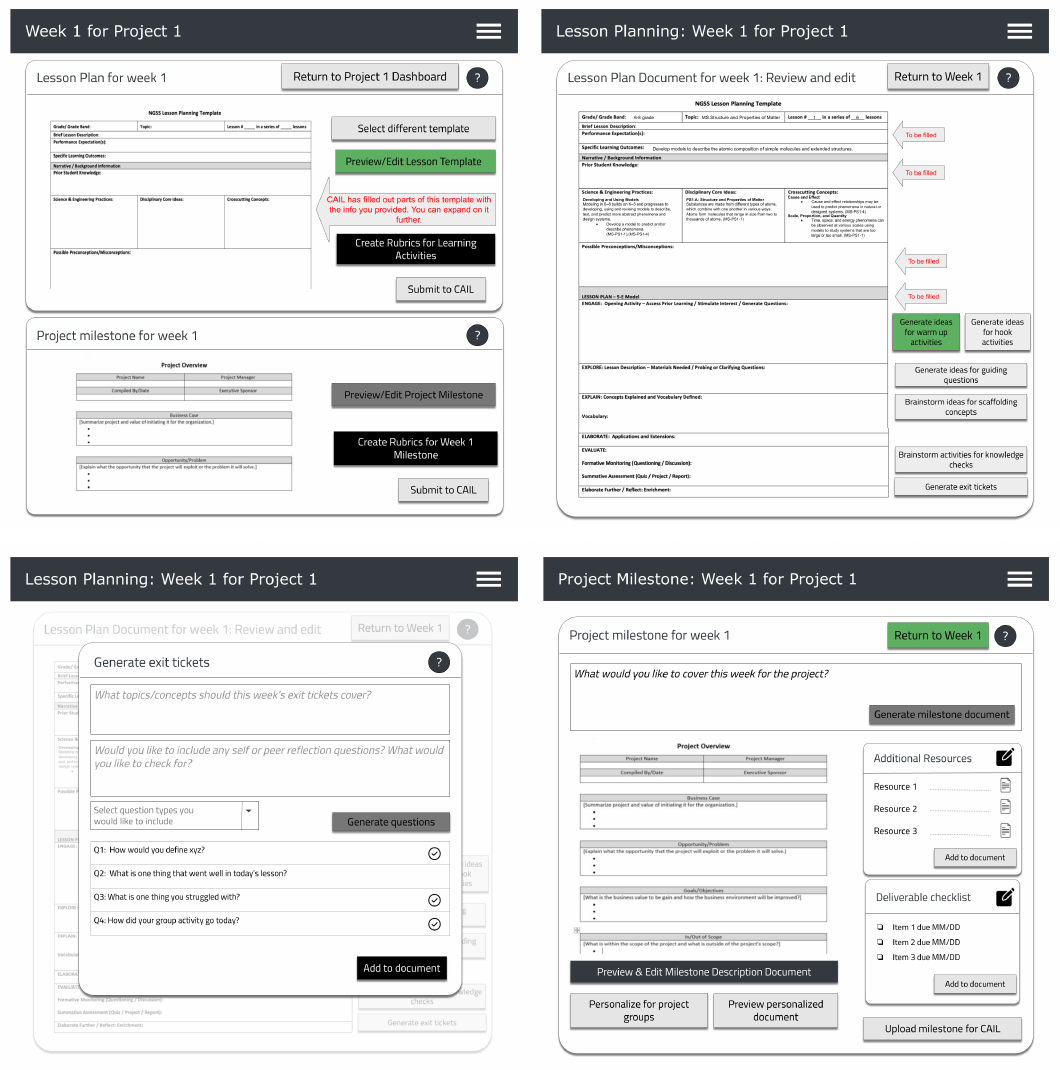}
    \caption{PBL LLM tool supports for lesson planning}
    \label{fig:lessons-wireframes}
    \Description{The figure shows wireframes for planning and organizing a project over multiple weeks. The "Project 1 Dashboard" provides an overview and weekly previews where teachers can plan lessons, assessments, scaffolding, and deliverables. Each week includes options to edit lesson plans, create rubrics, and submit details to an LLM tool (CAIL). Scaffolded LLM-powered lesson planning templates allow generating exit tickets, brainstorming activities, and creating project milestones with personalized documents, resources, and checklists to support learning objectives.}
\end{figure*}

\subsection{PBL GenAI Tool Wireframe Prototypes and Testing {(all Category themes)}}
From these discussions, we created wireframes for a teacher dashboard spanning the three categories of support (curriculum, assessments, and progress tracking). These wireframes explictly link their features to the core PBL challenges identified (from category B themes), and translate the identified needs (from category A themes) into actionable LLM-PBL solutions (in category C themes). \textbf{Below, we present these wireframes and the combined feedback received from teachers in both Study 1 (section 3.1.1) and Study 2 (section 3.2.1). 
By gathering perspectives from teachers with varying levels of PBL {expertise}, we wanted to ensure adaptability of LLM tool wireframes across diverse teaching methods and environments (addressing RO2).}
For brevity, we use the term CAIL (Collaborative AI for Learning) for the system and will refer to it as such throughout this section. 

\subsubsection{\textbf{Curriculum Supports}}
We focused on two key needs: brainstorming new project ideas with CAIL and co-implementing weekly or unit lesson plans based on these finalized ideas.

The project brainstorming wireframes (Figure \ref{fig:project-wireframe}) allow educators to expand on initial ideas gathered from classroom experiences, augmenting teacher creativity from the outset. Teachers input their learning goals and standards, ensuring \textbf{alignment with curricular objectives}. The tool integrates a curated list of U.S. national and state standards and allows teachers to \textbf{define expected project outputs} (e.g., slide decks, podcasts, portfolios), group size, project duration, and upload preferred lesson planning templates. CAIL offers multiple LLM-generated ideas, enabling teachers to refine and expand on them and provide students with \textbf{varied project options}.

\begin{figure*}
    \centering
    \includegraphics[width=1\linewidth]{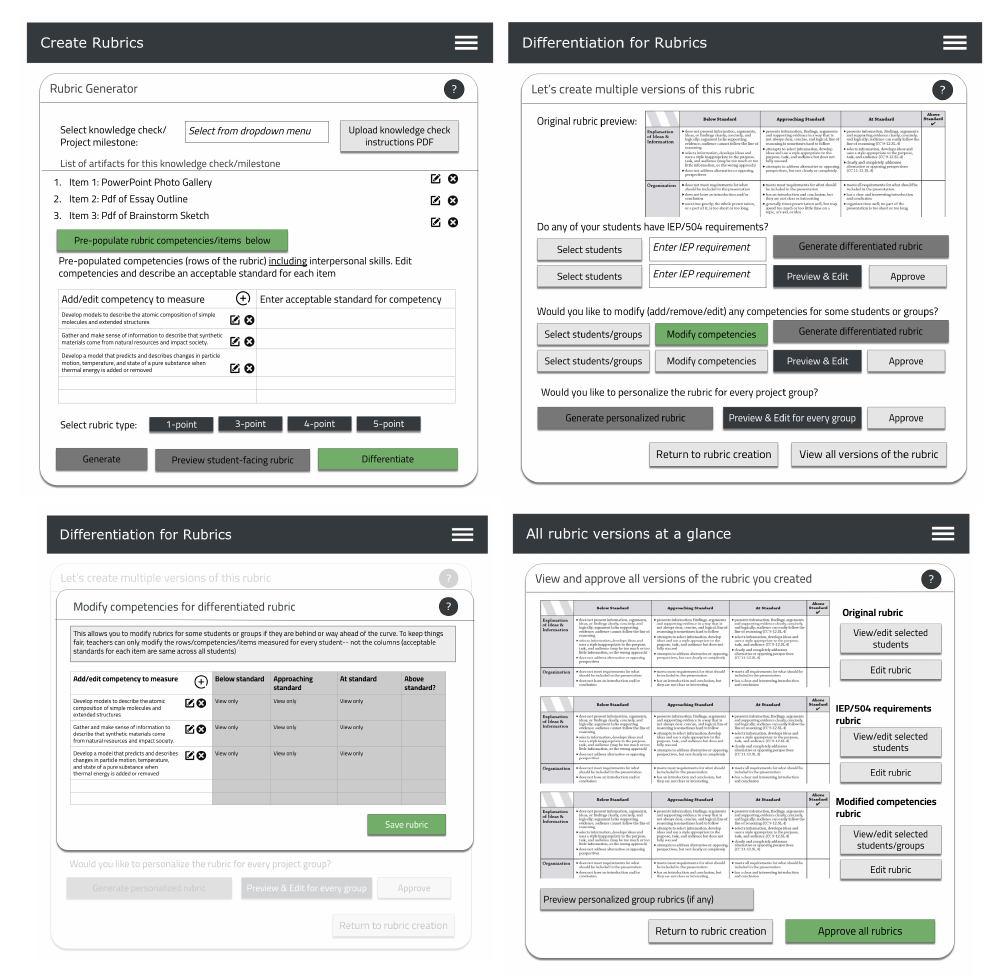}
    \caption{PBL LLM tool supports for rubric creation and differentiation}
    \label{fig:rubrics-wireframes}
    \Description{The figure shows wireframes for creating and differentiating rubrics for student assessments. The "Create Rubrics" screen allows teachers to select knowledge checks or project milestones, choose rubric types, and pre-populate competencies. The "Differentiation for Rubrics" section provides options to modify rubrics for different student needs, including those with IEP/504 requirements, and personalize rubrics for each group. The "All Rubric Versions at a Glance" screen lets teachers view, edit, and approve all rubric versions created, ensuring they meet diverse learning needs.}
\end{figure*}

Our participants suggested adding adjustable project durations, with the ability to \textbf{accommodate different class schedules}. Live collaboration features were also requested, enabling interdisciplinary teamwork among educators for formulating ideas. Teachers expressed interest in uploading custom standards and explored the idea of the LLM crosswalking standards to identify and address gaps. Combining multiple LLM-generated project ideas into a \textbf{choice board} for students was highlighted as a potential feature that could enhance differentiation. 

Teachers can then use CAIL for lesson planning (Figure \ref{fig:lessons-wireframes}) to implement selected project ideas. This includes creating formative assessments ("knowledge checks"), exit tickets, hooks, and scaffolded activities like games, discussions, peer learning, and problem-solving tasks. 
CAIL offers teachers a \textbf{structured lesson planning template} aligned with the standards they selected for their project ideas. 
For this wireframe, we picked an \href{https://www.nextgenscience.org/}{NGSS (Next Generation Science Standards)} example template that guides teachers through performance expectations, learning outcomes, and reflection elements. 
It provides targeted lesson plan suggestions, reducing planning time and offering guidance for both new and experienced PBL teachers. It maintains \textbf{institutional memory} by gathering data to inform future practices. Additionally, CAIL assists teachers in \textbf{creating milestone documents} and tailoring resources, such as checklists and schedules, to guide student groups through project phases.
Participants recognized the tool's long-term time-saving potential:

\begin{quote}
\textit{ “These are the 'money' tools for me. These buttons that help with generating these smaller activities that support the bigger project would be a huge time and brain saver.” --P01}
\end{quote}


However, they stressed the need for \textbf{thorough PD and microlearning modules} during rollout to help educators fully understand the tool's benefits, suggesting tutorials no longer than 20 minutes to avoid overwhelming them. 
Participants highlighted the importance of clear learning outcomes for each activity, aligning with competency-based teaching, and appreciated the iterative refinement of lesson plans. They also suggested that LLMs estimate activity time requirements while acknowledging their current limitations in precise time predictions.

\begin{figure}[ht]
    \centering
    \includegraphics[width=1\linewidth]{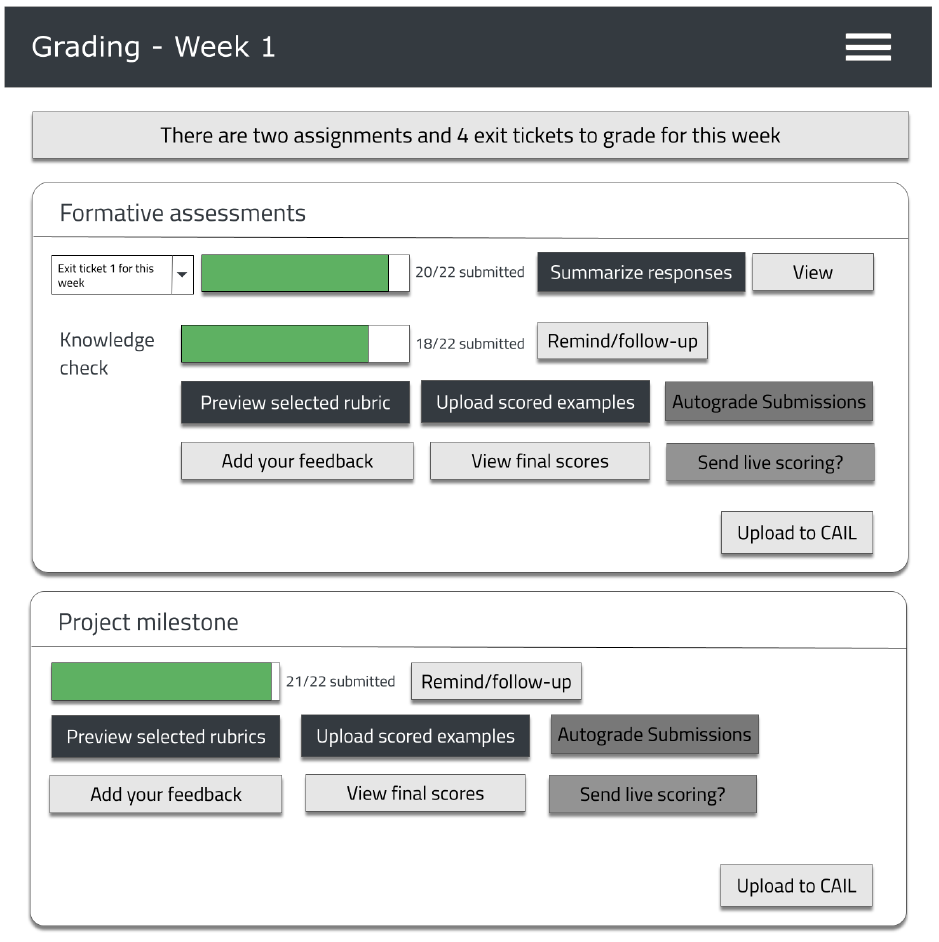}
    \caption{PBL LLM tool supports for grading}
    \label{fig:grading-wireframes}
    \Description{The figure shows wireframes for managing and grading formative assessments and project milestones. It displays submission statuses for assignments and exit tickets, with options to view responses, summarize feedback, remind or follow up with students, and autograde submissions. Teachers can preview rubrics, upload scored examples, provide feedback, and view final scores, with options to send live scoring updates.}
\end{figure}

\begin{figure*}[ht]
    \centering
    \includegraphics[width=1\linewidth]{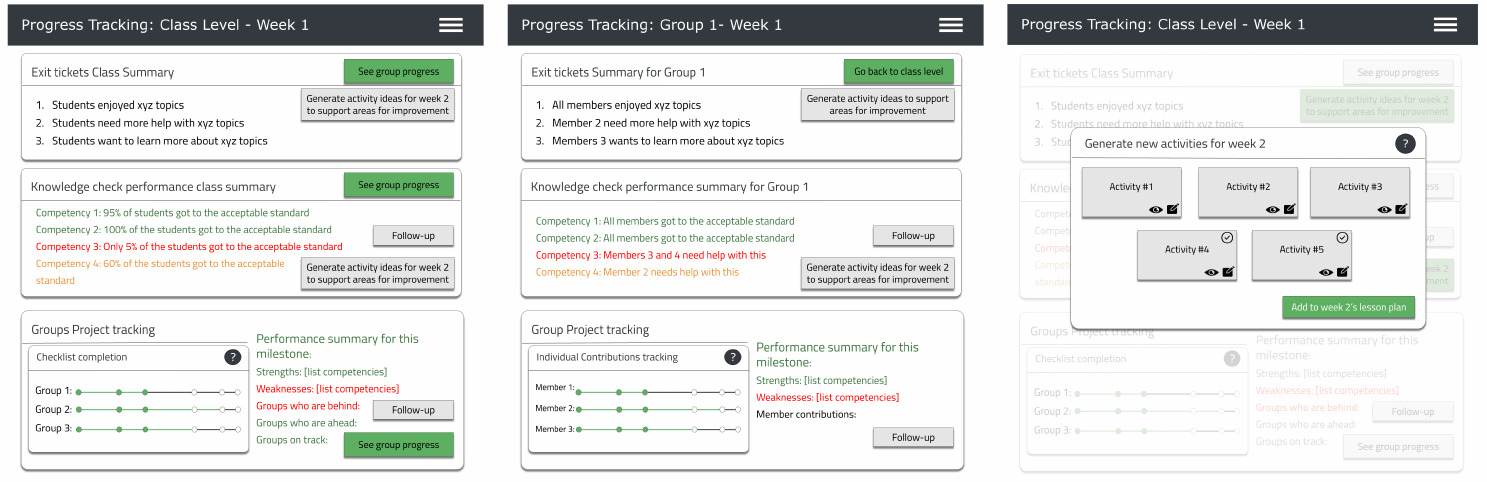}
    \caption{PBL LLM tool supports for progress tracking}
    \label{fig:progress-wireframes}
    \Description{The figure shows wireframes for tracking student progress at both class and group levels for Week 1. The "Progress Tracking: Class Level" screen provides summaries of exit ticket feedback, knowledge check performance, and group project tracking, highlighting areas where students excel or need improvement. The "Progress Tracking: Group 1" screen offers a detailed view of individual group performance, including contributions and areas needing support. There are options to generate new activities based on identified needs, follow up on competencies, and adjust lesson plans for the next week.}
\end{figure*}

\subsubsection{\textbf{Assessment Supports}} 
The assessment support section of the wireframes received mixed reactions from teachers, as discussed below. 

Teachers can use CAIL to \textbf{develop rubrics} (Figure \ref{fig:rubrics-wireframes}) for formative assignments and project milestones. They can input specific student artifacts for assessment and automatically populate rubric rows with competencies from the selected curriculum standards. Educators can further refine the criteria by editing acceptable standards for each rubric item, and \textbf{selecting a rubric type} (e.g., 1-point, 3-point, 4-point, or 5-point) to suit their classroom.
The tool also supports differentiation by generating multiple rubric versions tailored to individual student needs. Teachers can input \textbf{IEP/504 requirements} and pick competencies for specific students and groups, ensuring fair assessment by maintaining consistent standards while accommodating individual learning trajectories.

Participants suggested including examples to explain the different rubric types to help new teachers, \textbf{student work samples to set expectations}, and using visual highlights on rubrics to make feedback more accessible, noting that students often overlooked detailed feedback. 
Participants raised \textbf{privacy concerns} about using sensitive student IEP data, suggesting anonymized profiles and LMS integration for secure access. 
Participants emphasized the need for both \textbf{teacher- and student-facing rubrics}, with LLM assistance in simplifying language for students and using "I can" (\textit{e.g.,} "I can analyze data from a chart to make informed conclusions.") statements, to make these more empowering.

Figure \ref{fig:grading-wireframes} shows the CAIL grading page, which simplifies evaluating formative assessments and project milestones by tracking submissions, generating response summaries, and sending reminders for pending work. It allows teachers to upload scored examples as benchmarks, helping LLMs contextualize assessments. CAIL maps rubrics to relevant tasks, with LLM-generated scores complemented by required teacher feedback. An optional “live scoring” feature provides instant feedback to students if enabled by teachers.

Teachers were cautious about relying solely on LLMs for grading, particularly for subjective or creative work, emphasizing the \textbf{need for human oversight} to ensure feedback is personalized. Concerns were raised about biases in LLM grading, as well as its ability to handle atypical responses and physical artifacts that require nuanced understanding. Some participants like P05 (who taught PBL in online settings) did not want grading to be handled by the LLM at all:

\begin{quote}
\textit{“I understand why people want auto grading...but I think you learn so much about the individual students by interacting with their work. How else are you interacting with them? By seeing what they did and reading it. I wouldn't feel good about having AI grade it.” -- P05}
\end{quote}

Some teachers suggested using live scoring features to boost engagement and prompt revisions, \textbf{minimizing grading iterations}. However, others preferred the option to review and adjust LLM-generated comments before students received them and supported integrating custom feedback based on \textbf{classroom observations to complement LLM scores}.

\subsubsection{\textbf{Progress Tracking Supports}} 
The progress tracking section of the wireframes (Figure \ref{fig:progress-wireframes}) provides educators with an overview of student performance across individual, group, and classroom levels. The \textbf{Exit Tickets Summary} highlights topics students enjoy, areas needing help, and interests for future exploration, enabling targeted interventions; the Knowledge Check Performance Summary displays \textbf{class strengths and areas of concern} on competencies, allowing follow-up with struggling students; and the \textbf{Groups Project and Contributions} Tracking monitors group members' progress and individual contributions for effective interventions.

Participants recommended making progress data accessible to students individually and anonymously at the class level to enhance motivation and self-awareness. They emphasized \textbf{cross-disciplinary tracking for a holistic view of performance} and coordinated support across subjects. Although teachers found the progress data valuable for differentiation and scaffolding, they were concerned \textbf{about administrators using the data} \textbf{to evaluate their own performance.} They suggested creating an opt-in database to share projects and best practices, searchable using LLMs, based on student progress.



\section{DISCUSSION}

\subsection{Tensions in Designing PBL LLM Tools}
Our findings suggest that LLMs hold significant potential to enhance PBL practices by alleviating several teacher concerns. However, their use involves trade-offs that must be carefully managed for long-term effectiveness and sustainable use in the classroom. We present three tensions critical for the success of PBL-LLM tools: (1) technology choice and feasibility, (2) teacher agency, creativity, PD, and learning, and (3) balancing educational depth with reducing administrative workload.

\subsubsection{\textbf{Technology Choice and Feasibility:}} The selective use of LLMs is crucial to their effectiveness and sustainability in PBL. While LLMs can offer powerful capabilities in generating educational content and personalized learning materials \cite{owan2023exploring, lan2024teachers, zheng2024charting, weisz2024design}, teachers in our co-design workshops raised issues related to the accuracy and bias inherent in LLM-generated content, as seen in prior work \cite{zha2024designing, schneider2023towards, owan2023exploring}. LLMs could also 
undermine the fairness and inclusivity of educational practices due to this bias \cite{fang2024bias, rudolph2023chatgpt, schneider2023towards}. 
Our teachers expressed doubts about the reliability of LLMs in handling tasks that require a deep understanding of context or the subtle nuances involved in student expressions and creative multimodal artifacts \cite{rudolph2023chatgpt, schneider2023towards}. This skepticism reflects broader concerns about the limits of GenAI's interpretive abilities and risks associated with delegating too much of the evaluative process to technology \cite{zheng2024charting, rudolph2023chatgpt}. 

The feasibility of implementing LLMs in diverse educational settings was another significant issue. Our study highlighted concerns around logistical constraints, such as limited access to high-speed internet, adequate digital infrastructure, and computational resources, which can vary greatly across schools and districts \cite{zha2024designing}. These disparities raise concerns about digital equity, as schools with fewer resources may find themselves unable to leverage advanced LLM tools, potentially widening the gap in educational opportunities \cite{owan2023exploring, lan2024teachers}. Despite these challenges, teachers in our study were open to strategic and scoped integration of PBL-LLM tools by focusing on PBL-specific tasks best suited for LLM optimization to ensure their effectiveness in diverse contexts.

\subsubsection{\textbf{Teacher Agency, Creativity, Professional Development, and Learning: }}
Maintaining teacher agency and fostering creativity when integrating LLM tools were recurring areas of discussion in our work. 
Teachers expressed a strong preference for tools that align with their pedagogical goals, support creativity, and promote critical thinking among students \cite{lan2024teachers}. However, they also voiced concerns that excessive automation could diminish their role \cite{cope2021artificial}, particularly for novice PBL educators still learning the nuances of implementing high-quality PBL pedagogy. 

To address these concerns, our findings presented opportunities for LLM tools to offer teachers real-time guidance during resource creation while allowing for their active input.  
Rather than replacing teachers, the LLM could act as a co-educator and brainstorming partner, supporting their ability to design and implement activities \cite{nagy2023gen}. Our wireframes suggest opportunities for integrating LLM-driven PD modules into PBL tools, helping teachers build new skills and strategies.
Drawing on perspectives of teachers in our study, this training could also focus on helping PBL educators understand the tool's long-term benefits and customization options, ensuring it integrates smoothly into existing workflows. The in-tool LLM support could serve as an additional learning resource for novice PBL teachers by providing scaffolded guidance during lesson planning, aiding them in developing the skills needed to design and implement effective PBL lessons.
Teachers in our study were particularly enthusiastic about the prospect of learning prompt engineering and other LLM-specific skills, seeing these as new ways to enhance their creativity and adapt technology to their classroom needs \cite{so2024enhancing}. 
To realize these benefits, the design of LLM tools must involve ongoing collaboration among developers, teachers, and PD designers. 

\subsubsection{\textbf{Balancing Educational Depth with Reducing Administrative Workload:}}
LLM tools have the potential to significantly reduce administrative burdens in PBL settings, such as managing group projects, tracking student progress, and providing individualized feedback \cite{chan2023ai} that  
can free up teachers to focus on more meaningful instructional activities \cite{samala2024unveiling}. However, our results caution against allowing efficiency gains to oversimplify the complex aspects of PBL, such as developing students' interpersonal skills, cultivating creativity, and supporting critical thinking—areas that require nuanced, human-centered guidance 
\cite{lan2024teachers}. Similar to previous work, our teachers also feared that excessive automation could diminish opportunities for spontaneous, in-the-moment learning experiences and the development of soft skills that are crucial for student growth \cite{chen2024artificial, hutson2023rethinking}. There is thus a pressing need to identify which administrative tasks can be effectively automated without compromising the integrity of the learning experience. 

\subsection{LLM Design Recommendations for Project-Based Learning}

\begin{figure}
    \centering
    \includegraphics[width=1\linewidth]{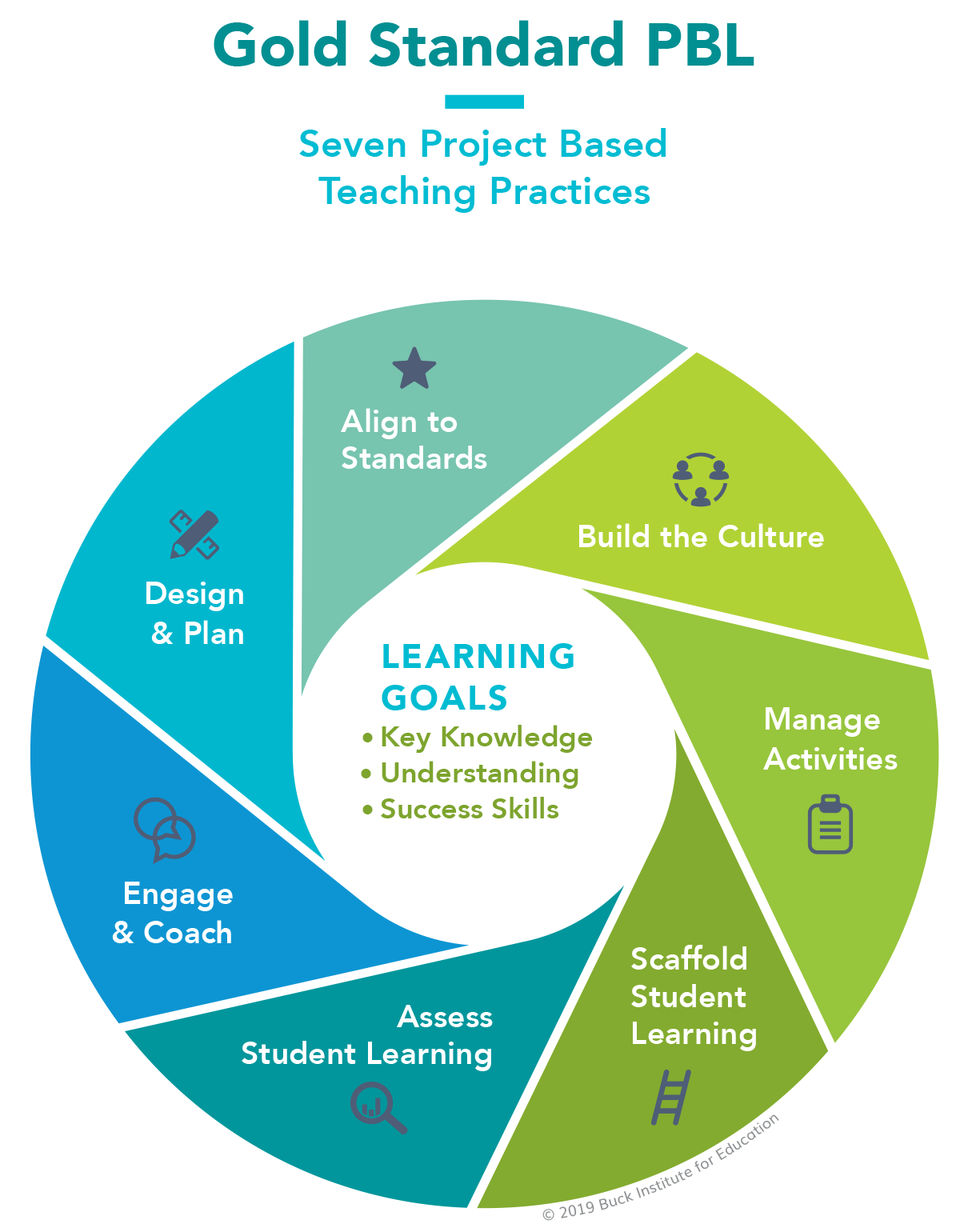}
    \caption{Gold Standard Framework: Project Based Teaching Practices}
    \label{fig:gold-pbl}
    \Description{The figure illustrates the "Gold Standard PBL" framework, which outlines seven project-based teaching practices. At the center are the core learning goals: key knowledge, understanding, and success skills. The surrounding practices include: "Design & Plan," "Align to Standards," "Build the Culture," "Manage Activities," "Scaffold Student Learning," "Assess Student Learning," and "Engage & Coach." Each practice contributes to achieving the central learning goals in a project-based learning environment.}
\end{figure}

Keeping the above tensions in mind, we 
present design recommendations for educational technology designers and educators when integrating LLMs into PBL (addressing RO3).
We draw on the \textbf{“Gold Standard PBL: Project Based Teaching Practices” framework from the Buck Institute for Education} that guides teachers in implementing high-quality PBL \cite{PBLWorks}. 
We briefly describe the framework's seven PBL teaching practices 
(Figure \ref{fig:gold-pbl}) and unpack potential ways LLM systems can mitigate implementation barriers. 
\textbf{We focus on recommendations that alleviate teachers' administrative workload in implementing high-quality PBL, allowing them to dedicate more time to the creative and fulfilling aspects of teaching.}
We advocate for these discussions to actively include new teachers, curriculum coaches, school boards, admins, information technology (IT) and special education committees — stakeholders our teachers emphasized as vital but often underrepresented in education tool design \cite{pnevmatikos2020stakeholders}.
We structure the two sub-sections below by mapping a combination of these practices to the wireframe supports we developed and the design recommendations they informed.

\subsubsection{\textbf{Recommendations for Project Ideation, Curriculum Planning, and Management}}- \\

\noindent\fbox{
    \parbox{\columnwidth}{%
\textbf{Gold Standard PBL Teaching Practice \#1: Design and Plan: }\textbf{“}\textit{Teachers create or adapt a project for their context and students, and plan its implementation from launch to culmination while allowing for some degree of student voice and choice.”}}%
}
\noindent\fbox{
    \parbox{\columnwidth}{%
\textbf{Gold Standard PBL Teaching Practice \#2: Align to Standards:} \textit{\textbf{“}}\textit{Teachers use standards to plan the project and make sure it addresses key knowledge and understanding from subject areas to be included.” }
}%
}
\noindent\fbox{
    \parbox{\columnwidth}{%
\textbf{Gold Standard PBL Teaching Practice \#3: Manage Activities:} \textit{\textbf{“}}\textit{Teachers work with students to organize tasks and schedules, set checkpoints and deadlines, find and use resources, create products and make them public.”}}%
}
\noindent\fbox{
    \parbox{\columnwidth}{%
\textbf{Gold Standard PBL Teaching Practice \#4: Scaffold Student Learning:}\textit{ “Teachers employ a variety of lessons, tools, and instructional strategies to support all students in reaching project goals.”}
}%
}

Based on the feedback we received on project ideation and curriculum planning supports, we outline the following design considerations for Practices 1, 2, 3, and 4: \\

\textbf{Recommendation 1: Augment Teacher Creativity with LLM Tools}

\textit{Design LLM tools to augment and elevate teacher creativity by empowering educators to lead the project brainstorming process from the outset while leveraging LLMs to support mundane, peripheral tasks. Build on established pedagogical practices of PBL educators, reducing the need for creating resources from scratch.}

Our results showed evidence of PBL teachers grappling with balancing the depth of knowledge required for specific projects with the breadth of content mandated by the curriculum. Teachers also raised concerns about the excessive contextual information required by current LLM systems and their ability to effectively prompt the LLM for specific tasks. A proposed solution is structured input: teachers specify contextual details such as specific learning goals, standards, desired final project artifacts, and duration of implementation, and LLM tools generate ideas aligned with curricular objectives and tailored to classroom needs. This structured input approach not only supports experienced educators but also provides novice PBL teachers with foundational knowledge for constructing high-quality PBL units. \\

\textbf{Recommendation 2: Provide Teacher-Directed Scaffolding with Flexibility}

\textit{Design LLM tools for PBL that intentionally prioritize and mandate core teacher inputs, while offering optional customization features.}

This recommendation is rooted in two ideas from our research:
(1) the need to promote high-quality pedagogy by ensuring that key LLM functionalities are integrated into teaching practices, while also providing flexibility for educators to choose supplementary features that support their unique classroom needs,
(2) ensuring the LLM system has sufficient contextual information from the required inputs to generate high-quality resources.
For example, requiring teachers to input or select educational standards for their project ideas should be mandatory, ensuring alignment with curriculum requirements and learning outcomes \cite{kurtz2024strategies}. In contrast, optional features can accommodate varying school contexts, as our results showed that teachers sometimes chose not to use extended features depending on their specific needs. \\

\textbf{Recommendation 3: Ensure Flexibility in Scheduling for Diverse Classroom Needs}

\textit{Design LLM tools to accommodate specific temporal structures of diverse educational environments.}

This includes accounting for various class schedules, including block scheduling (a method of dividing time into distinct blocks and assigning each block to a specific task or activity) \cite{chen2019revisiting, savery2015overview}. Since methods for chunking projects varied vastly across teachers of different grade levels and disciplines in our workshops’ data \cite{kokotsaki2016project}, features to parameterize project duration based on hours, days, or weeks are necessary to provide the flexibility teachers require when brainstorming ideas for projects.\\

\textbf{Recommendation 4: Promote Differentiation with LLM-Generated Project Ideas}

\textit{Design LLM tools to generate multiple project ideas that teachers can combine into choice boards, thereby enhancing differentiation and student agency.}

\begin{figure*}
    \centering
    \includegraphics[width=0.9\linewidth]{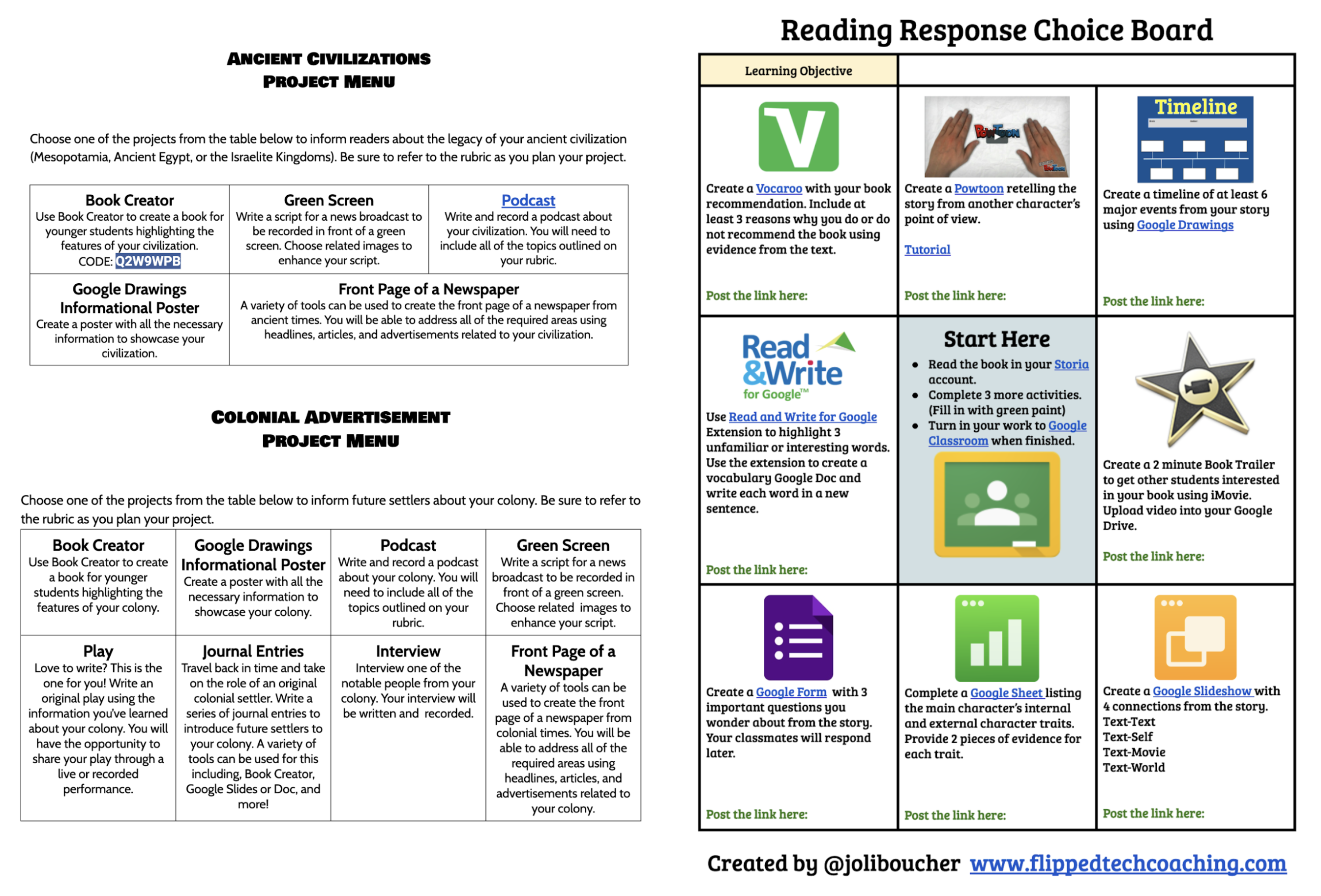}
    \caption{Examples of choice boards used by our teachers to promote student agency}
    \label{fig:choice-boards}
    \Description{The figure features three example choice boards for projects. The "Ancient Civilizations Project Menu" allows students to choose projects such as creating a book, writing a news script, recording a podcast, or designing a newspaper front page to showcase the legacy of ancient civilizations. The "Colonial Advertisement Project Menu" offers similar creative options like developing posters, journal entries, interviews, and plays to engage students in learning about colonial history. The "Reading Response Choice Board" provides diverse activities for students to demonstrate their understanding of a book, including creating videos, timelines, book trailers, and interactive documents using various digital tools.}
\end{figure*}

Implementing choice boards with multiple project options was a common practice among our teachers (Figure \ref{fig:choice-boards}), but proved challenging due to time constraints and difficulties with inventing novel, creative ideas each time. To address this, the tool should allow for continuous modification and regeneration of LLM-generated content, enabling teachers to refine and align projects with their educational goals and classroom needs.\\

\textbf{Recommendation 5: Streamline Lesson Planning with LLM-powered Standards Templates}

\textit{Embed LLM supports within existing lesson planning templates from common educational standards to streamline the process of generating engaging instructional content.} 

This can reduce the time and cognitive load for teachers when aligning PBL activities with academic standards. Consistent with other scholars, our research revealed that teachers often struggled with the complexity of designing knowledge-check activities, driving questions, and other project components while ensuring alignment with standards \cite{savery2015overview, ertmer2006jumping}. The LLM system could provide immediate, context-sensitive suggestions and resources, tailored to specific lesson goals and standards, allowing teachers to focus on customizing their lesson plans rather than building them from scratch. This feature is especially beneficial for less experienced teachers, easing their adoption of PBL principles, while also providing experienced educators with tools to document and share their expertise.

\subsubsection{\textbf{Recommendations for Progress Tracking and Assessments}}- \\

\noindent\fbox{
    \parbox{\columnwidth}{%
\textbf{Gold Standard PBL Teaching Practice \#5: Build the Culture: }\textit{“Teachers explicitly and implicitly promote student independence and growth, open-ended inquiry, team spirit, and attention to quality.”}
}%
}

\noindent\fbox{
    \parbox{\columnwidth}{%
\textbf{Gold Standard PBL Teaching Practice \#6: Assess Student Learning: }\textit{“Teachers use formative and summative assessments of knowledge, understanding, and success skills, and include self and peer assessment of team and individual work.”}
}%
}

\noindent\fbox{
    \parbox{\columnwidth}{%
\textbf{Gold Standard PBL Teaching Practice \#7: Engage and Coach: }“\textit{Teachers engage students in their learning and work alongside them to identify when they need skill-building, redirection, encouragement, and celebration.”}
}%
}
Based on the feedback we received for our assessments and progress tracking supports, we outline the following design considerations for Practices 5, 6, and 7:\\

\textbf{Recommendation 6: Develop Equitable and Actionable Rubrics}

\textit{Incorporate a default rubric format within the LLM tool that promotes fair and equitable grading practices, aligning with high-quality PBL pedagogy.}

Our research data revealed the importance of having educators collaborate with LLMs to set actionable and easily understandable expectations within rubrics that meet required competencies. The LLM tool should also facilitate the creation of two separate rubric sets: one for teachers and administrators, focusing on competencies from established standards, and another for students, emphasizing clear, actionable expectations tied to project outputs.
\begin{itemize}
    \item \textbf{Recommendation 6.1: }\textbf{Implement single-point rubrics as the default rubric type within LLM tools for PBL.}
Single-point rubrics use a single column for feedback instead of multiple performance levels. Research shows that such rubrics are quicker and easier for teachers to create since they don't need to anticipate all possible ways students might \textit{“fail expectations”} \cite{gonzalez_single-point_2015}. For students, these rubrics are simpler to understand, focusing on clear target expectations. They also promote higher-quality feedback, as teachers highlight specific strengths and areas for improvement, fostering a growth mindset and iterative learning.

    \item \textbf{Recommendation 6.2: }\textbf{Augment rubric items with detailed teacher feedback.}
Rubric items must be augmented with detailed feedback from teachers. Designers of the LLM tools should explore the possibility of providing a starting point for feedback from student submissions, allowing teachers to elaborate and personalize their notes \cite{takale2024assessing, han2024teachers}.

    \item \textbf{Recommendation 6.3: }\textbf{Integrate LLM-assisted positive feedback to complement constructive criticism alongside rubrics.}
Our teachers mentioned struggling to provide granular positive feedback due to time constraints. Praise and celebration, bolstered by relevant specifics, can instill student confidence and promote balanced assessments.\\
\end{itemize}

\textbf{Recommendation 7: Ensure Privacy in LLM-Driven Differentiation}

\textit{Ensure that data processing and LLM interactions for differentiating student resources occur locally on the teacher's device or within a secure school network.}

Teachers in our study expressed significant concerns about privacy when using LLMs to differentiate rubrics for students with IEPs, particularly regarding the sharing of sensitive student information. To address these concerns, LLM tools should either operate locally or be integrated securely within existing Learning Management Systems (LMS) as an app or extension \cite{wang2024artificial, zha2024designing}. This approach ensures alignment with privacy regulations such as COPPA (Children's Online Privacy Protection Rule). Pseudonyms can also automatically replace identifying student information before being used for differentiation. \\

\begin{figure*}
    \centering
    \includegraphics[width=0.79\linewidth]{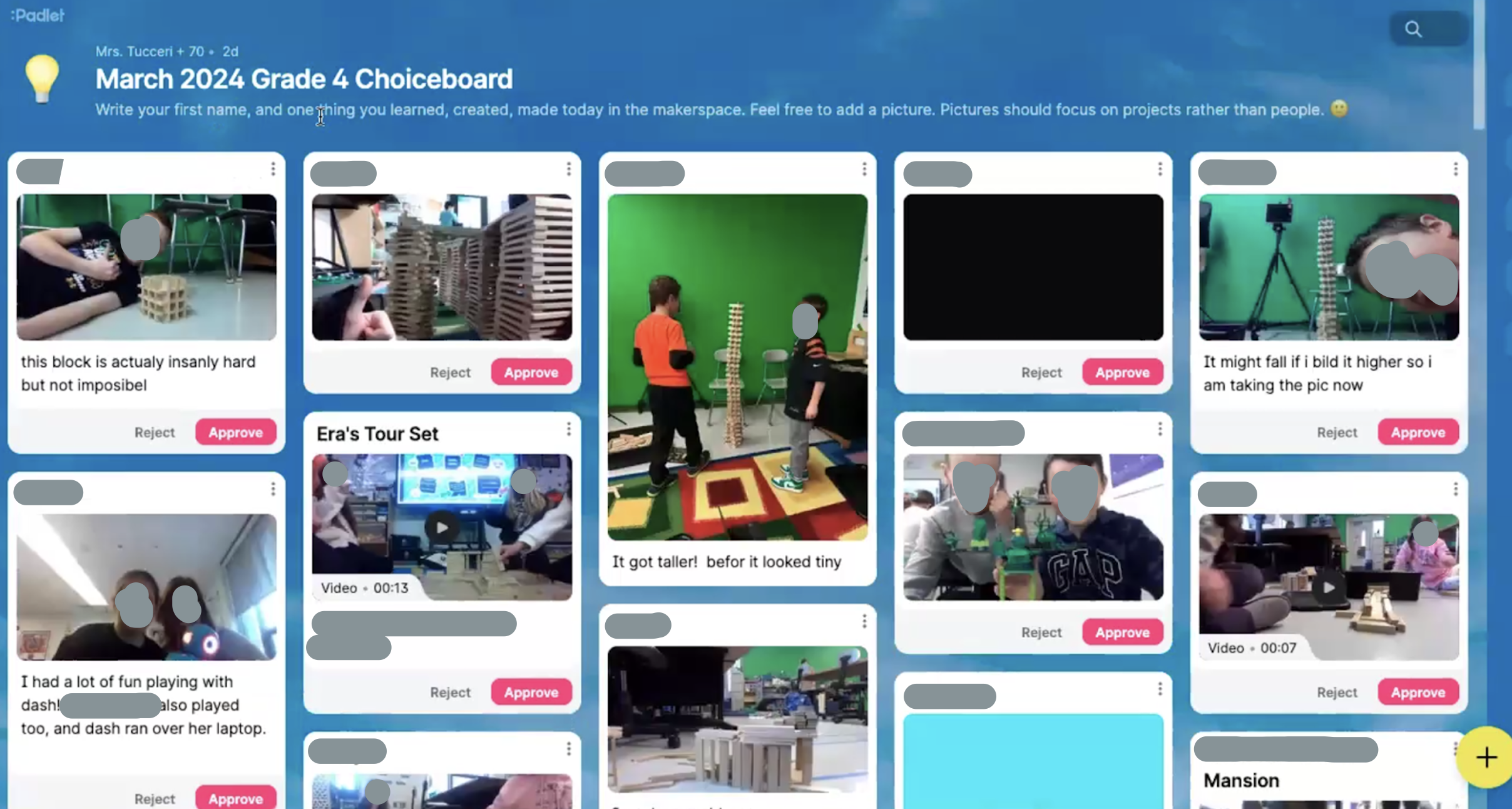}
    \caption{Examples of Padlets used by our teachers to document student progress and self-reflections}
    \label{fig:padlet}
    \Description{The figure shows a Padlet board featuring various student posts showcasing their projects and activities completed in the makerspace. Each post includes a photo or video with a caption describing the project, such as building structures, playing with educational robots, and creating sets. There are options for the teacher to approve or reject each submission, encouraging students to share their learning experiences.}
\end{figure*}

\textbf{Recommendation 8: Integrate Project Management Scaffolds and Supports}

\begin{itemize}
    \item \textbf{Recommendation 8.1: PBL LLM tools should include student-facing project management features that complement teacher supports}. Teachers in our study found balancing open-ended problem-solving with personalized guidance challenging, especially in managing group projects. The LLM can serve as a brainstorming partner and offer scaffolding to guide students, providing support when teachers are unavailable \cite{kim2020bot, kim2020bot, kim2024engaged, chase2009teachable, tan2022systematic, shaer2024ai}. LLM conversational agents could potentially capture key discussion points and synthesize them for review, acting as self-check tools to ensure groups complete daily tasks and take appropriate next steps \cite{lan2024teachers, tanga2024exploration, cai2024advancing}. Additionally, these agents can analyze and monitor participation frequency, and assess the quality of contributions to ensure equitable participation within groups \cite{ouyang2023integration, sekeroglu2019student, tan2022systematic}.

    \item \textbf{Recommendation 8.2: Design LLM
    tools that assist teachers in providing personalized feedback during active class time.} By alerting teachers to groups needing immediate attention and tracking group progress, LLMs can help prioritize the teacher’s time and provide notes for asynchronous feedback \cite{dutta2024enhancing, kanchon2024enhancing}. This support allows teachers to efficiently manage multiple groups without overwhelming their workload, ensuring that all student groups receive necessary guidance.

    \item \textbf{Recommendation 8.3: Implement robust LLM-powered portfolio supports that enable students to submit evidence of their processes and reflections, documenting their learning journey from ideation to project completion}. 
    Prior research has underscored the importance of processes (\textit{how} students did things throughout the project, from ideation to prototyping, for example, including challenges faced and \textit{what} they did about them), and reflection (growth or trajectory over time) \cite{fields2023communicating}. 
    Our teachers used Padlets for students to submit their daily work (e.g., snapshots, videos, responses) and as a reflective tool for exit tickets (Figure \ref{fig:padlet}).
    LLMs can support the creation of individual portfolios to showcase personal skills and group portfolios to document project development \cite{lin2023portfolio}. Additionally, LLMs should incorporate self and peer reflection activities, as well as conflict resolution tools, drawn from toolkits like \href{https://makered.org/beyondrubrics/toolkit/}{“Beyond Rubrics”}, to ensure meaningful, ongoing assessment \cite{mlpblbrief, kokotsaki2016project}. By embedding these elements into portfolios, LLM can facilitate both formative and summative assessments, offering a comprehensive view of student learning.

    \item \textbf{Recommendation 8.4: Design LLM tools to support rather than replace teacher involvement in the grading process by organizing and presenting data that tracks student progress}. Our study revealed that teachers highly value their agency in the grading process as it enables them to deeply understand student thinking, provide personalized feedback, and make informed decisions. Teachers also expressed concerns about the quality of LLM-generated feedback and the subjectivity involved in grading PBL student multimodal artifacts. By organizing data on milestones, monitoring group performance, and curating exit ticket responses, LLM tools can alleviate administrative burdens and provide a structured starting point for teachers. This ensures that teachers remain at the center of the grading process, using LLMs as a means to enhance their ability to deliver nuanced and targeted feedback, ultimately leading to more effective and personalized student assessments.
\end{itemize}

\section{LIMITATIONS AND FUTURE WORK}

Our study has limitations that suggest avenues for future research. First, our focus on PBL in the context of the United States limits the generalizability of the findings; future research should explore the applicability of LLM-driven PBL tools in culturally diverse educational settings globally. {We also acknowledge the small sample size and absence of collected racial demographics information as limitations for generalizability.} Additionally, our study involved a self-selected group of teachers who were likely more interested in GenAI and PBL, potentially biasing the findings toward a more positive outlook. Future work could include longer-term studies that investigate how such tools might engage teachers who are more resistant to PBL or GenAI integration. The wireframes developed in this study are only a starting point;  building and testing a working product through iterative design cycles is crucial to evaluate the feasibility of GenAI for the proposed features. Future research should focus on co-designing student-facing supports that align with teacher tools, fostering an integrated flow of information and feedback to create a balanced ecosystem within the LLM-driven PBL experience. 

\begin{acks}
We would like to thank Robert Parks and Raechel Walker for their invaluable assistance with the workshops. We also sincerely appreciate pivotal input received from Christina Bosch during key stages of the study. Finally, this study would have been impossible without the 41 remarkable educators, all deeply passionate about project-based learning, who brought their enthusiasm
to every step of the process.
\end{acks}

\bibliographystyle{ACM-Reference-Format}
\bibliography{sample-base}










\end{document}